\documentclass[12pt]{article}

\usepackage{amsmath,mathtools,amsfonts}
\usepackage{amssymb}
\usepackage{mathrsfs}
\usepackage{bm}
\usepackage[dvips]{graphicx}
\usepackage[all]{xy}
\usepackage{cite}

\newcommand{\llangle}{\langle\!\langle}
\newcommand{\rrangle}{\rangle\!\rangle}
\newcommand{\bd}[1]{\boldsymbol{#1}}
\newcommand{\id}{\mathbb{I}}

\newcommand{\NSNS}{NS\textrm{-}NS}
\newcommand{\RNS}{R\textrm{-}NS}
\newcommand{\NSR}{NS\textrm{-}R}
\newcommand{\RR}{R\textrm{-}R}

\begin{document}

\baselineskip=17pt

\begin{titlepage}
\rightline{\tt YITP-20-154}
\rightline\today
\begin{center}
\vskip 2.5cm
 {\Large \bf {
Tree-level S-matrix of superstring field theory
with homotopy algebra structure
}}
\vskip 1.0cm
{\large {Hiroshi Kunitomo}}
\vskip 1.0cm
{\it {Center for Gravitational Physics}}, 
{\it {Yukawa Institute for Theoretical Physics}}\\
{\it {Kyoto University}},
{\it {Kyoto 606-8502, Japan}}\\
kunitomo@yukawa.kyoto-u.ac.jp

\vskip 2.0cm

{\bf Abstract}
\end{center}
\noindent
We show that the tree-level S-matrices of the superstring field
theories based on the homotopy-algebra structure agree with those obtained 
in the first-quantized formulation.
The proof is given in detail for the heterotic string field theory.
The extensions to the type II and open superstring
field theories are straightforward.

\end{titlepage}

\tableofcontents

\section{Introduction}

After the pioneering work by Witten \cite{Witten:1986qs},  
research on the superstring field theory was falling into a long period of 
stagnation, except for a few important developments 
\cite{Preitschopf:1989fc,Arefeva:1989cp,Berkovits:1995ab,Okawa:2004ii,Berkovits:2004xh}. 
Recently, however, several important progress has been made one after another
\cite{Jurco:2013qra,Iimori:2013kha,Erler:2013xta,Erler:2014eba,Erler:2015lya,Erler:2015rra,
Sen:2015hha,Sen:2015uaa,Konopka:2016grr,Kunitomo:2015usa,Erler:2016ybs,
Matsunaga:2016zsu,Goto:2016ckh,Erler:2016rxg,Kunitomo:2016kwh,Erler:2017onq,
Matsunaga:2014wpa,Kunitomo:2019glq,Kunitomo:2019kwk}, 
and several complete superstring field theories are now established.

Now, there are three complementary formulations, each of which has advantages and disadvantages:
the Wess-Zumino-Witten (WZW) -like formulation, 
the formulation based on the homotopy-algebra structure,
and the formulation accompanied with an extra free field.
The WZW-like formulation was first proposed for the open superstring by Berkovits 
in his ingenious paper \cite{Berkovits:1995ab} and afterward extended to the heterotic 
string field theory \cite{Okawa:2004ii,Berkovits:2004xh}.
Although both of them were originally limited to the NS sector, 
but have recently been extended to
a complete form including the Ramond sector \cite{Kunitomo:2015usa,Kunitomo:2019glq}.
Attempts to construct a WZW-like action for the type II superstring field theory 
are also being made \cite{Matsunaga:2016zsu,Kunitomo:2019kwk}.
The homotopy-algebra-based formulation, 
the open superstring field theory with an $A_\infty$ structure \cite{Erler:2013xta},
and the heterotic and type II superstring field theories with an $L_\infty$ structure 
\cite{Erler:2014eba}, was pioneered by the Munich group. 
Also in this formulation, all the constructions were initially
limited to the NS or NS-NS sector, but soon have been extended to those including 
the Ramond sector, and now completed \cite{Erler:2016ybs,Kunitomo:2019glq,Kunitomo:2019kwk}.
The formulation accompanied with an extra free field has been developed by Sen for the heterotic and
type II superstring field theories \cite{Sen:2015hha,Sen:2015uaa}.
In this formulation, a pair of Ramond string fields are introduced, which double the degrees of freedom 
but half of them is cleverly
decoupled from the physical world as a free field.
It has been shown that the formulation can also apply to the open superstring 
field theory \cite{Konopka:2016grr}. 

In this paper, we consider the homotopy-algebra-based superstring field theories
and show that their tree-level physical S-matrices agree with those calculated
by the first-quantized method \cite{Friedan:1985ge,Witten:2012bh}.
In the bosonic string theory, the amplitude 
for each process is given by the integration over the moduli space of punctured Riemann
surface. The string field theory provides a triangulation of the moduli space, 
each region of which is filled with the contribution from a Feynman diagram.
Each contribution is necessary to be connected smoothly at the boundaries
so that the sum is an integral over the entire moduli space.
%
It is well known that this requirement is essentially equivalent to
requiring the action to be gauge invariant
\cite{Giddings:1986wp,Saadi:1989tb,Kugo:1989aa,Zwiebach:1992ie}.
For the superstring field theories, on the other hand, the amplitudes are given by
the integral over the super moduli space, so we must also take into account
the contribution of the odd moduli integration.
It can be incorporated by insertions of the picture changing operator (PCO) 
\cite{Witten:2012bh}, which apparently seems to disturb
the smooth connection between contributions from Feynman diagrams. 
However, the gauge-invariant action constructed by utilizing the homotopy 
algebra structure also includes terms that may fill the gap as contributions 
from the vertical integration \cite{Sen:2015hia}.
As the result, the superstring field theories 
with homotopy algebra structure reproduce the S-matrices obtained 
in the first-quantized formulation, which the purpose of this paper
is to prove.

The study along this direction was previously performed for several tree-level 
four-string amplitudes: four-NS string amplitude in 
the open superstring\cite{Erler:2013xta}, and arbitrary four-string amplitudes
in the heterotic string field theory \cite{Kunitomo:2019glq}, and some typical 
four-string amplitudes in the type II superstring field theory 
\cite{Kunitomo:2019kwk}.~\footnote{The explicit confirmations for four-NS 
\cite{Berkovits:1999bs}
and arbitrary four- and five-string amplitudes \cite{Kunitomo:2016bhc}
were also performed in the WZW-like formulation. 
These can also be considered as a confirmation of those 
in the homotopy-algebra-based formulation
since the two theories are related to a simple field redefinition.}
In addition to these confirmations by explicit calculation,
a general proof for the tree-level S-matrix was also given based on 
the minimal model with the homological perturbation theory (HPT) \cite{Konopka:2015tta}.
This proof is, however, still restricted to the amplitudes with 
the external NS(-NS) strings, and the proof for the complete S-matrix 
is still lacking.
We extend it proof to the general S-matrix and complete the proof 
in the homotopy-algebra-based superstring field theories. 

The paper is organized as follows.
In section \ref{S hetero}, we give a general proof for the heterotic 
string in detail. We first summarize some basic properties of the heterotic 
string field theory with cyclic $L_\infty$ structure in section \ref{hetero summ}, 
focusing on what is needed for the proof.
Then, in section \ref{hetero S gen}, we give the 
S-matrix generating function at tree-level 
in a closed form based on a general argument given in 
Ref.~\cite{Jevicki:1987ax,Arefeva:1974jv}. The result agrees with that
given by using the argument based on the (almost) minimal model with HPT. 
Using the S-matrix generating function,
we show that the tree-level physical S-matrix agrees with that calculated
using the first-quantized method in section \ref{hetero eval}. 
The proof is given by generalizing
the method used in Ref.~\cite{Konopka:2015tta}.
It is straightforward to extend the proof to the type II superstring
field theory, which is given in section \ref{type II}.
After summarizing the basic properties in section \ref{type II summ},
we give a proof for type II superstring field theory in section \ref{type II eval}.
The extension to the open superstring field theory is also straightforward
if we generalize the construction method in Ref.~\cite{Erler:2016ybs} to the one
applicable more general $A_\infty$ structure following the way given
in Ref.~\cite{Kunitomo:2019glq}. 
Starting from giving
such a generalized open superstring field theory 
in section \ref{open summ}, we prove the agreement of
the S-matrix in section \ref{open eval}. Section \ref{summary} is
devoted to the summary and discussion. Finally, six appendices are included. 
Appendix \ref{app S matrix} is devoted to deriving
the final form of the S-matrix generating functional, which is not given
in the text since some technical details are required.
In Appendix \ref{app 1}, we explain 
how the S-matrix is also obtained
as the (almost) minimal model via HPT.
Some proofs to confirm the consistency of the perturbation are also given.
Appendices \ref{app 2} and \ref{app 3} 
are devoted to proving some formulas used in the text. In Appendix \ref{app der},
we show that the generalized open superstring field theory given in Section \ref{open} 
reduces to that in Ref.~\cite{Erler:2016ybs} as a special case based on the Witten's 
vertex. 
A simple proof for the cyclicity of the $A_\infty$ structure in the generalized
theory is also given in Appendix \ref{app cyclicity}, which was not so simple
in the previous construction.

\section{Tree-level S-matrix of heterotic string field theory}\label{S hetero}

Let us first take the heterotic string field theory 
and show that it reproduces the tree-level S-matrix calculated by 
the first-quantized method.

\subsection{Heterotic string field theory with cyclic 
$L_\infty$ structure}\label{hetero summ}

We start by briefly summarizing how the heterotic string field theory based on the
$L_\infty$ structure is constructed.
The first-quantized heterotic string theory is obtained by combining
the left-moving (holomorphic) superconformal field theory 
and the right-moving (anti-holomorphic) bosonic conformal field theory.
The former consists of a matter sector with $c=15$\,, the fermionic (conformal) ghosts $(b,c)$\,, 
and the bosonic (superconformal) ghosts $(\beta,\gamma)$\,.
The latter consists of a matter sector with $c=26$\,, and 
the fermionic ghosts $(\overline{b},\overline{c})$. 
It is useful to \lq bosonize\rq\, the bosonic ghosts $(\beta,\gamma)$ to
a pair of fermions $(\eta,\xi)$ and a chiral boson $\phi$\,. The Hilbert space
of the bosonized ghosts $(\eta,\xi;\phi)$
is called the large Hilbert space $\mathcal{H}_l$\,.
In contrast, the Hilbert space 
of the bosonic ghosts $(\beta,\gamma)$
is called the small Hilbert space $\mathcal{H}_{s}$, 
which can be embedded in $\mathcal{H}_l$ as
states $\Phi$ satisfying $\eta\Phi=0$ with a fixed picture number.\footnote{
For notational simplicity, the zero-mode of $\eta$-ghost 
appearing frequently
is simply denoted as $\eta$\,.}
The subspaces with different picture numbers provide equivalent representations
of $\mathcal{H}_s$\,, and 
can be related to each other by one-to-one mapping using PCO.
The heterotic string field takes a value in $\mathcal{H}_s$ as described below 
in detail.

The heterotic string field $\Phi$ is Grassmann even and satisfying
the closed string constraints:
\begin{equation}
 b_0^-\Phi\ =\ L_0^-\Phi\ =\ 0\,.
\label{restrict closed}
\end{equation}
It has ghost number 2 and two components
\begin{equation}
 \Phi\ =\ \Phi_{NS} + \Phi_R\ \in\ 
\mathcal{H}^{res}\ =\ \mathcal{H}_{NS} + \mathcal{H}_R^{res}\,,
\end{equation}
where $\mathcal{H}_{NS}$ ($\mathcal{H}_R$) is the small Hilbert space of 
the NS (Ramond) sector with picture number $-1$ ($-1/2$).
The Ramond Hilbert space $\mathcal{H}_R^{res}$ is 
further restricted by an extra condition:
\begin{equation}
\mathcal{H}_R^{res}\ =\ \{\Phi_R\in\mathcal{H}_R\mid XY\Phi_R\ =\ \Phi_R \}\,,
\end{equation}
where $X$ and $Y$ are defined by
\begin{equation}
 X\ =\ -\delta(\beta_0)G_0+\delta'(\beta_0)b_0\,,\qquad
 Y\ =\ -2c_0^+\delta'(\gamma_0)\,.
\end{equation}
The operator $X$ is the PCO on states 
with picture number $-3/2$\,, and commutative with the BRST operator 
$Q$\,. In the large Hilbert space, it can be written as the BRST exact form
$X=\{Q,\Xi\}$ with 
\begin{equation}
 \Xi\ =\ \xi_0 + (\Theta(\beta_0)\eta\xi_0-\xi_0)\Pi_{-3/2}
+(\xi_0\eta\Theta(\beta_0)-\xi_0)\Pi_{-1/2}\,.
\end{equation}
Here, $\Pi_n$ is the projection operator onto the states with picture number $n$\,.
The operator $Y$ acts on states with picture number $-1/2$ as
an inverse of $X$ in the sense that it satisfies
\begin{equation}
 XYX\ =\ X\,,\qquad YXY\ =\ Y\,.
\label{XYX}
\end{equation}
We can show that $XY$ used to define $\mathcal{H}^{res}$
is a projection operator acting on the sates with picture number $-1/2$. 
The ghost number of the string field is equal to the basis states
for the classical string field.

For later use, it is useful to introduce a notation
\begin{equation}
 \mathcal{G}\ =\ \pi^0+X\pi^1\,,\qquad
 \mathcal{G}^{-1}\ =\ \pi^0+Y\pi^1\,,
\end{equation}
with the projection operator $\pi^0$ $(\pi^1)$ onto
the NS (Ramond) component.
Then, the restricted Hilbert space 
$\mathcal{H}^{res}$
can concisely be written as
\begin{equation}
\mathcal{H}^{res}\ =\ \{\Phi\in\mathcal{H}_s=\mathcal{H}_{NS}+\mathcal{H}_R
\mid\mathcal{G}\mathcal{G}^{-1}\Phi\ =\ \Phi\}\,.
\label{restricted}
\end{equation}
The restricted Hilbert space $\mathcal{H}^{res}$ 
is closed under the action of the BRST operator: $XYQXY=QXY$\,,
and any state $\mathcal{B}\in\mathcal{H}^{res}$ can be expanded in the ghost
zero-mode as
\begin{align}
 \mathcal{B} =&\ \mathcal{B}_{NS} + \mathcal{B}_{R}
\nonumber\\
=&\
\left(b_{NS}-c_0^+B_{NS}\right) 
+ \left(b_R - \frac{1}{2}(\gamma_0+2c_0^+G)B_R\right)\,,
\label{zero mode form}
\end{align}
where $G=G_0+2\gamma_0b_0$\,. 
In particular, we denote the string field $\Phi\in\mathcal{H}^{res}$ as
\begin{equation}
 \Phi\ =\ \left(\phi_{NS}-c_0^+\psi_{NS}\right) 
+ \left(\phi_R - \frac{1}{2}(\gamma_0+2c_0^+G)\psi_R\right)\,.
\label{0-mode exp}
\end{equation}

We can define
three symplectic forms for the large, small and restricted Hilbert space, 
$\omega_l$\,, $\omega_s$ and $\Omega$ by
\begin{subequations} \label{symplectic}
 \begin{alignat}{4}
 \omega_l(\Phi_1,\Phi_2)\ =&\ (-1)^{|\Phi_1|}{}_l\langle \Phi_1|c_0^-| \Phi_2\rangle_l\,,
\qquad& \Phi_1,\Phi_2\in&\ \mathcal{H}_l\,,\label{symp l}\\
 \omega_s(\Phi_1,\Phi_2)\ =&\ (-1)^{|\Phi_1|}{}_s\langle \Phi_1|c_0^-|\Phi_2\rangle_s\,,
\qquad& \Phi_1,\Phi_2\in&\ \mathcal{H}_s\,,\label{symp s}\\
 \Omega(\Phi_1,\Phi_2)\ =&\ 
(-1)^{|\Phi_1|}{}_s\langle \Phi_1|c_0^-\mathcal{G}^{-1}|\Phi_2\rangle_s\,,\qquad& 
\Phi_1,\Phi_2\in&\ \mathcal{H}^{res}\,.\label{symp res}
\end{alignat}
\end{subequations}
For $\Phi_1,\Phi_2\in\mathcal{H}^{res}$\,, 
they are related as
\begin{equation}
\Omega(\Phi_1\,,\Phi_2)\ =\ \omega_s(\Phi_1\,,\mathcal{G}^{-1}\Phi_2)\
=\ \omega_l(\xi_0\Phi_1,\mathcal{G}^{-1}\Phi_2)\,.
\end{equation}
%
%
We also use their bilinear map representation 
defined by
\begin{equation}
 \begin{alignedat}{6}
\langle\omega_l|\,:\,
\mathcal{H}_l&\otimes\mathcal{H}_l\qquad & 
&\ {\longrightarrow}&\qquad &\mathbb{C}
\\
&\ \rotatebox{90}{$\in$} & & &  &\rotatebox{90}{$\in$}
\\
 \Phi_1&\otimes\Phi_2\qquad 
& &\longmapsto&\qquad \omega_l(\Phi&_1\,,\Phi_2)\,,
\end{alignedat}
\end{equation} 
and 
\begin{equation}
 \langle\omega_s|\ =\ \langle\omega_l|(\xi_0\otimes\id)\,,\qquad
 \langle\Omega|\ =\ \langle\omega_l|(\xi_0\otimes\mathcal{G}^{-1})\,.
\end{equation}
Note that the natural inner product in $\mathcal{H}^{res}$ defined by $\Omega$
has the off-diagonal form
\begin{equation}
 \Omega(\Phi_1,\Phi_2)\ =\ \llangle\phi_{NS1}|\psi_{NS2}\rrangle 
+ \llangle\psi_{NS1}|\phi_{NS2}\rrangle 
+\llangle\phi_{R1}|\psi_{R2}\rrangle + \llangle\psi_{R1}|\phi_{R2}\rrangle \,,
\label{inner pro}
\end{equation}
after integrating out the ghost zero-modes.\footnote{
The double bra-ket represents the BPZ inner-product 
after integrating out the ghost zero-modes \cite{Kunitomo:2019kwk}.}

The action of heterotic string field theory with cyclic $L_\infty$ structure
is written as
\begin{equation}
 I[\Phi]\ =\ \sum_{n=0}^\infty\frac{1}{(n+2)!}\,
\Omega(\Phi\,,L_{n+1}(\underbrace{\Phi\,,\cdots\,,\Phi}_{n+1}))\,,
\label{small action}
\end{equation}
which is invariant under the gauge transformation
\begin{equation}
 \delta\Phi\ =\ \sum_{n=0}^\infty\frac{1}{n!}\, 
L_{n+1}(\underbrace{\Phi\,,\cdots\,,\Phi}_n\,,\Lambda)\,.
\label{small gauge tf}
\end{equation}
Here, $L_1=Q$ and the multi-string products $L_{n+2}(\Phi_1,\cdots,\Phi_{n+2})$
are graded commutative, which satisfy the $L_\infty$ relations
\begin{align}
\sum_\sigma\sum_{m=1}^n(-1)^{\epsilon{(\sigma)}}
\frac{1}{m!(n-m)!}
L_{n-m+1}(L_m(\Phi_{\sigma(1)},\cdots,\Phi_{\sigma(m)}), 
\Phi_{\sigma(m+1)},\cdots,\Phi_{\sigma(n)})\ =\ 0\,,
\label{L infinity}
\end{align}
and the cyclicity condition
\begin{equation}
 \Omega(\Phi_1\,, L_n(\Phi_2\,,\cdots\,,\Phi_{n+1}))\ =\ 
- (-1)^{|\Phi_1|}\,\Omega(L_n(\Phi_1\,,\cdots\,,\Phi_n)\,, \Phi_{n+1})\,.
\label{cyclicity}
\end{equation}
The symbol $\sigma$ in (\ref{L infinity})
denotes the permutation from $\{1,\cdots,n\}$
to $\{\sigma(1),\cdots,\sigma(n)\}$ and the factor
$\epsilon(\sigma)$ is the sign factor of permutation of string fields from
$\{\Phi_1,\cdots,\Phi_n\}$ to $\{\Phi_{\sigma(1)},\cdots,\Phi_{\sigma(n)}\}$\,.
The set of the string products satisfying these conditions is called
a cyclic $L_\infty$ algebra $(\mathcal{H}^{res},\Omega,\{L_n\})$.

For constructing the string products of the cyclic $L_\infty$ algebra, 
we use the coalgebra representation, which is, for example, summarized
in \cite{Erler:2015uba,Kunitomo:2019glq}.
The heterotic string products are represented by the (Grassmann) odd
coderivation $\bd{L}=\bd{Q}+\bd{L}_{int}$ 
acting on the symmetrized tensor algebra 
$\mathcal{SH}^{res}=\oplus_{n=0}^\infty(\mathcal{H}^{res})^{\wedge n}$\,.
However, it is not very good idea to directly construct the cyclic $L_\infty$
algebra $(\mathcal{H}^{res},\Omega,\bd{L})$ since $\Omega$ is asymmetric between 
the NS and Ramond sectors. 
Instead, we considered, in \cite{Kunitomo:2019glq}, an cyclic $L_\infty$ algebra 
$(\mathcal{H}_l,\omega_l,\bd{Q}-\bd{\eta}+\bd{B})$ first.
The coderivation $\bd{B}$ acting on $\mathcal{SH}_l
=\oplus_{n=0}^\infty(\mathcal{H}_l)^{\wedge n}$ is constructed so that
an extended generating function, 
\begin{equation}
 \bd{B}(s,t)\ =\ 
\sum_{m,n,r=0}^\infty s^m t^n \bd{B}^{(n)}_{m+n+r+1}|^{2r}\,,
\label{B in st}
\end{equation}
related to $\bd{B}$ as $\bd{B}=\bd{B}(0,1)$\,,
satisfies two differential equations
\begin{subequations}\label{diff eq} 
 \begin{align}
 \partial_t\bd{B}(s,t)\ =&\ [\bd{Q},\bd{\lambda}(s,t)]
+ [\bd{B}(s,t),\bd{\lambda}(s,t)]^1
  + s\, [\bd{B}(s,t),\bd{\lambda}(s,t)]^2\,,\\
\partial_s\bd{B}(s,t)\ =&\ [\bd{\eta},\bd{\lambda}(s,t)]
- t\, [\bd{B}(s,t),\bd{\lambda}(s,t)]^2\,,
\end{align}
\end{subequations}
which simultaneously determine another even coderivation
\begin{equation}
 \bd{\lambda}(s,t)\ =\ \sum_{m,n,r=0}^\infty s^m t^n \bd{\lambda}^{(n+1)}_{m+n+r+2}|^{2r}\,,
\end{equation}
called the (generating function of) gauge products.
Here, the brackets $[\,\cdot\,,\cdot\,]^1$ and $[\,\cdot\,,\,\cdot\,]^2$
are defined from the (graded) commutators by projecting
the cyclic Ramond number as
\begin{equation}
 [\bd{A}|^{2r},\bd{B}|^{2s}]^1\ =\ [\bd{A}|^{2r},\bd{B}|^{2s}]|^{2(r+s)}\,,\qquad
 [\bd{A}|^{2r},\bd{B}|^{2s}]^2\ =\ [\bd{A}|^{2r},\bd{B}|^{2s}]|^{2(r+s-1)}\,.
\label{com 1 2}
\end{equation}
For the coderivation $\bd{B}(s,t)$ constructed to satisfy (\ref{diff eq}), 
the relations
\begin{subequations}\label{extended L infty} 
\begin{align}
&\ [\bd{Q},\bd{B}(s,t)] + \frac{1}{2}\,[\bd{B}(s,t),\bd{B}(s,t)]^1
+ \frac{s}{2}\,[\bd{B}(s,t),\bd{B}(s,t)]^2\ =\ 0\,,\\
&\ [\bd{\eta},\bd{B}(s,t)] - \frac{t}{2}\,[\bd{B}(s,t),\bd{B}(s,t)]^2\ =\ 0\,,
\end{align}
\end{subequations}
hold, which reduce at $(s,t)=(0,1)$
to the $L_\infty$ relation 
for the coderivation $\bd{Q}-\bd{\eta}+\bd{B}$\,. 
We have shown that 
the coderivation $\bd{L}$ 
representing the heterotic string products can be obtained
from $\bd{B}$ as
\begin{subequations} 
 \begin{align}
\bd{L}\ =&\ \hat{\bd{F}}^{-1}(\bd{Q}+\pi^0\bd{B})\hat{\bd{F}}\,,\\
 \pi_1\bd{L}_{int}\ =&\ \mathcal{G}\pi_1\bd{l}\,,\qquad
 \pi_1\bd{l}\ =\ \pi_1\bd{B}\hat{\bd{F}}\,.
\label{L int}
 \end{align}
\end{subequations}
Here, $\pi_n$ is the projection operator 
\begin{equation}
\pi_n\,:\, \mathcal{SH}^{res}\rightarrow(\mathcal{H}^{res})^{\wedge n}\,.
\end{equation}
The invertible cohomomorphism $\hat{\bd{F}}^{-1}$ is defined by
\begin{equation}
 \pi_1\hat{\bd{F}}^{-1}\ =\ \pi_1\id - \Xi\pi_1^1\bd{B}\,,
\label{hat F inv}
\end{equation}
with $\pi_1^1=\pi^1\pi_1$ ($\pi^0_1=\pi^0\pi_1$), 
and $\id\in\mathcal{SH}^{res}$
is the (multiplicative) identity of the symmetrized tensor algebra. 
%
%
Using a property of cohomomorphisms
\begin{equation}
 \hat{\bd{F}}\ =\ \pi_0 + \sum_{n=1}^\infty\frac{1}{n!}\left(\pi_1\hat{\bd{F}}\right)^{\wedge n}\,,
\end{equation}
and the relation
\begin{equation}
 \pi_1\hat{\bd{F}}\ =\ \pi_1\id + \Xi\pi_1^1\bd{l}\,,
\label{hat F}
\end{equation}
obtained by acting $\hat{\bd{F}}$ from the right of both sides of (\ref{hat F inv}), 
we can rewrite the second equation in (\ref{L int}) as
a self-consistent equation, which we will use later:
%
\begin{equation}
 \pi_1\bd{l}\ =\ \sum_{n=0}^\infty \pi_1\bd{B}_{n+2}\left(\frac{1}{(n+2)!}
\Big(\pi_1\id + \Xi\pi_1^1\bd{l}\Big)^{\wedge(n+2)}\right)\,.
\label{recursive l}
\end{equation}

Finally, we note that 
the bracket $[\,,]^{1}$ or $[\,,]^{2}$ 
can also be defined by projecting the intermediate state onto 
the NS or Ramond state, respectively. For example, for coderivations 
$\bd{A}_{n+1}$ and $\bd{B}_{m+1}$\,, we find that
\begin{align}
 [\bd{A}_{n+1},\bd{B}_{m+1}]^1\ =&\ 
\bd{A}_{n+1}\big(\pi^0_1\bd{B}_{m+1}\wedge\id_{n}\big)
-(-1)^{AB}\bd{B}_{m+1}\big(\pi^0_1\bd{A}_{n+1}\wedge\id_m\big)\,,\\
 [\bd{A}_{n+1},\bd{B}_{m+1}]^2\ =&\ 
\bd{A}_{n+1}\big(\pi^1_1\bd{B}_{m+1}\wedge\id_{n}\big)
-(-1)^{AB}\bd{B}_{m+1}\big(\pi^1_1\bd{A}_{n+1}\wedge\id_m\big)\,.
\end{align}
In this definition, it is easy to see that $[\,,]=[\,,]^1+[\,,]^2$ holds. 
We can rewrite the relations in Eqs.~(\ref{diff eq}) and
(\ref{extended L infty}) using this representation as
\begin{subequations}\label{diff eq fixed ex}
 \begin{align}
 \partial_t\bd{B}_{n+2}(s,t)\ =&\
[\bd{Q},\bd{\lambda}_{n+2}(s,t)]
+\sum_{m=0}^{n-1}\Big(
\bd{B}_{m+2}(s,t)\big(\pi(s)\pi_1\bd{\lambda}_{n-m+1}(s,t)\wedge\id_{m+1}\big)
\nonumber\\
&\hspace{45mm}
-\bd{\lambda}_{m+2}(s,t)\big(\pi(s)\pi_1\bd{B}_{n-m+1}(s,t)\wedge\id_{m+1}\big)
\Big)\,,
\label{diff t alt}\\
 \partial_s\bd{B}_{n+2}(s,t)\ =&\
[\bd{\eta},\bd{\lambda}_{n+2}(s,t)]
-\sum_{m=0}^{n-1}\Big(
\bd{B}_{m+2}(s,t)\big(t\pi^1_1\bd{\lambda}_{n-m+1}(s,t)\wedge\id_{m+1}\big)
\nonumber\\
&\hspace{45mm}
-\bd{\lambda}_{m+2}(s,t)\big(t\pi^1_1\bd{B}_{n-m+1}(s,t)\wedge\id_{m+1}\big)
\Big)\,,
\label{diff s alt}
\end{align}
\end{subequations}
and
\begin{subequations}\label{L infty fixed ex}
 \begin{align}
[\bd{Q},\bd{B}_{n+2}(s,t)]\ =&\ 
  -\sum_{m=0}^{n-1}\bd{B}_{m+2}(s,t)
\big(\pi(s)\pi_1\bd{B}_{n-m+1}(s,t)\wedge\id_{m+1}\big)\,,
\label{Q B alt form}\\
[\bd{\eta},\bd{B}_{n+2}(s,t)]\ =&\ 
  \sum_{m=0}^{n-1}\bd{B}_{m+2}(s,t)
\big(t\pi^1_1\bd{B}_{n-m+1}(s,t)\wedge\id_{m+1}\big)\,,
\label{Q eta alt form}
\end{align}
\end{subequations}
with $\pi_1\bd{B}_{n+2}(s,t)=\pi_1\bd{B}(s,t)\pi_{n+2}$\,,
$\pi_1\bd{\lambda}_{n+2}(s,t)=\pi_1\bd{\lambda}(s,t)\pi_{n+2}$
and $\pi(s)=\pi^0+s\pi^1$\,.
%
We will use these forms of the relations later.

\subsection{S-matrix generating function}\label{hetero S gen}

In the quantum field theory,
a perturbative amplitude is conventionally calculated 
utilizing the Feynman's method: 
drawing the possible (connected and amputated) Feynman diagrams 
for a specific process, combining the propagators and vertices, 
and evaluating them with integrating the loop momentum (momenta) 
if it is the one at the loop level.
Also for the open \cite{Erler:2013xta}, heterotic \cite{Kunitomo:2019glq}, and 
type II \cite{Kunitomo:2019kwk} superstring field theories, 
we have already calculated various tree-level four-point amplitudes based
on the homotopy-algebra-based formulation.
\footnote{Several four- and five-point amplitudes
have also been calculated by the WZW-like open superstring field theory
\cite{Berkovits:1999bs,Kunitomo:2016bhc}. } 
In this section, we discuss all the (tree-level) amplitudes collectively
by considering the S-matrix generating functional at the tree level.

We first quantize the theory following the BV formalism.
From the off-diagonal form of the inner product (\ref{inner pro}) in $\mathcal{H}^{res}$\,,
we can identify $\phi=\phi_{NS}+\phi_R$ and $\psi=\psi_{NS}+\psi_R$ in
(\ref{0-mode exp}) to the field and anti-field of the BV formalism 
in the gauge-fixed basis \cite{Kohriki:2012pp}. 
In this basis, the classical BV master action 
has the same form as the classical action 
(\ref{small action}), 
\begin{equation}
 I[\Phi]\ =\ \sum_{n=0}^\infty\frac{1}{(n+2)!}\,
\Omega(\Phi\,,L_{n+1}(\underbrace{\Phi\,,\cdots\,,\Phi}_{n+1}))\,,
\label{master action}
\end{equation}
but $\Phi$ is now the quantum string field including the states in $\mathcal{H}^{res}$ with all
the ghost numbers.\footnote{We use the same symbol for the classical and quantum
string fields and their Hilbert spaces, but it would not be confused.
The ghost number of the quantum field remains two, which is
defined by the sum of two kinds of ghost numbers:
the ghost number of basis states and the ghost number of coefficient space-time fields
(space-time ghost number).}
The Siegel-Ramond (SR) gauge is obtained by simply setting $\psi=0$\,.
We call the Hilbert space of the quantum field in this gauge
$\mathcal{H}_{SR}$\,: $\Phi=\phi\in \mathcal{H}_{SR}$\,.
Then the gauge fixed action
\begin{equation}
 I[\phi]
=\ \frac{1}{2}\langle\phi|c_0^-\mathcal{G}^{-1}Q|\phi\rangle
+ \sum_{n=3}^\infty\frac{1}{n!}\langle\phi|c_0^-
|l_{n-1}\left(\phi^{n-1}\right)\rangle\,,
\label{gf action}
\end{equation}
is invariant under the BRST transformation\footnote{
This is a schematic expression. 
See Ref.~\cite{Zwiebach:1992ie}
for precise treatment.}
\begin{equation}
 \delta_{BRST}\phi\ =\ \left.\frac{\partial I[\Phi]}{\partial \psi}\right|_{\psi=0}\,.
\end{equation}
From (\ref{gf action}), we can read off 
the propagator 
\begin{subequations}\label{Feynman}
\begin{equation}
\Pi\ \equiv\ - \frac{b_0^+b_0^-\mathcal{G}}{L_0^+}
\int^{2\pi}_0\frac{d\theta}{2\pi}e^{-i\theta L_0^-}\,,
\label{prop hetero}
\end{equation}
and the $n$-point vertices  
\begin{equation}
 c_0^-l_{n-1}\,,\label{vert hetero}
\end{equation}
\end{subequations}
for the Feynman rules to calculate amplitudes.
Here, however, instead of considering each amplitude independently,
we consider the generating functional of S-matrix elements which 
can be obtained by evaluating the effective action at its stationary 
configuration \cite{Jevicki:1987ax}.
At the tree-level, in particular, it reduces to 
the classical action evaluated at the classical solution and is calculated
as follows.
%
%
%

Let us first define the \textit{on-shell} subspace, 
\begin{equation}
 \mathcal{H}_0\ =\ 
\{\Phi\in\mathcal{H}^{res}\,
\mid L_0^+\Phi_{NS}=G_0\Phi_R=0\}\,,
\end{equation}
and the BRST invariant projection operator,
\begin{equation}
 \qquad
P_0\,:\, \mathcal{H}^{res} \longrightarrow \mathcal{H}_0\,,\qquad
 P_0^2\ =\ P_0\,, \quad [Q,P_0]\ =\ 0\,.
\end{equation}
Then, we introduce an operator
\begin{equation}
 Q^+\ =\ \frac{1}{L_0^+}b_0^+(1-P_0)\,,
\end{equation}
satisfying
\begin{subequations} 
\begin{align}
&\ Q^+Q + QQ^+ + P_0\ =\ \id\,,\\
&\ Q^+P_0\ =\ P_0Q^+\ =\ Q^+Q^+\ =\ 0\,,\label{prop Q+}
\end{align}
\end{subequations}
which defines a Hodge-Kodaira decomposition of $\mathcal{H}^{res}$\,,
\begin{equation}
\mathcal{H}^{res}\ =\ \mathcal{H}^p+\mathcal{H}^t+\mathcal{H}^u\,,
\label{HK}\\
\end{equation}
with
\begin{equation}
\mathcal{H}^p\ =\ P_0\mathcal{H}^{res}\,,\qquad
\mathcal{H}^t\ =\ QQ^+\mathcal{H}^{res}\,,\qquad
\mathcal{H}^u\ =\ Q^+Q\mathcal{H}^{res}\,.
\end{equation}
This $Q^+$ is called a (contracting) homotopy operator of $Q$\,, which is
compatible with $\Omega$\,: 
\begin{equation}
 \Omega(\mathcal{H}_0,\mathcal{H}^u)\ =\ \Omega(\mathcal{H}^u,\mathcal{H}^u)\ =\ 0\,.
\end{equation}
%
Under the SR gauge condition $\psi=0$\,, $\mathcal{H}^t=\emptyset$ and
the quantum field $\phi$ is decomposed
to the on-shell and off-shell subspaces:
\begin{equation}
 \phi\ \in\ \mathcal{H}_0+\overline{\mathcal{H}}_0\,,
\end{equation}
with
\begin{equation}
 \mathcal{H}_0\ =\ \mathcal{H}^p\cap\mathcal{H}_{SR}\ =\ P_0\mathcal{H}_{SR}\,,\qquad
 \overline{\mathcal{H}}_0\ =\ \mathcal{H}^u\cap\mathcal{H}_{SR}\ =\
(1-P_0)\mathcal{H}_{SR}\,.
\end{equation}
The classical equation of motion in the SR gauge
can now be written in the form of an integral equation:
\begin{equation}
\phi\ =\ \phi_0 - Q^+\pi_1\bd{L}_{int}(e^{\wedge\phi})\,,\qquad
\phi_0\in\mathcal{H}_0\,.
\label{recursion 1}
\end{equation}
At the tree level, the S-matrix generating functional is 
given by evaluating the classical action at the solution 
of the equation (\ref{recursion 1}). Its explicit expression can be found 
using the coalgebra representation as follows.

Define the homotopy, projection and identity operators, $\bd{H}$\,, $\hat{\bd{P}}$\,,
and $\hat{\bd{I}}$ by
\begin{subequations}\label{homotopy ops} 
 \begin{align}
 \bd{H}\ =&\ \sum_{r,s=0}^\infty\frac{1}{(r+s+1)!}\,(-Q^+)\wedge
\underbrace{\id_1\wedge\cdots\wedge\id_1}_r\wedge\underbrace{
P_0\wedge\cdots\wedge P_0}_s\,,
\label{def H}\\
\hat{\bd{P}}\ =&\ \sum_{n=0}^\infty P_n\ =\
\sum_{n=0}^\infty\frac{1}{n!}\underbrace{P_0\wedge\cdots\wedge P_0}_n\,,
\label{def P}\\
\hat{\bd{I}}\ =&\ \sum_{n=0}^\infty \id_n\ =\
\sum_{n=0}^\infty\frac{1}{n!}\underbrace{\id_1\wedge\cdots\wedge \id_1}_n\,,
\label{def I}
\end{align}
\end{subequations}
which act on the symmetric tensor algebra $\mathcal{SH}_{SR}$ 
and satisfy the relations
\begin{subequations}\label{basic prop}
 \begin{align}
&\ \hat{\bd{P}}\ =\ \hat{\bd{I}} +
\bd{H}\bd{Q} + \bd{Q}\bd{H}\,,
\label{prop 1}\\
&\ \hat{\bd{P}}^2\ =\ \hat{\bd{P}}\,,\qquad
[\bd{Q},\hat{\bd{P}}]\ =\ 0\,,\qquad
\label{prop 0}\\
&\ \bd{H}\hat{\bd{P}}\ =\ 
\hat{\bd{P}}\bd{H}\ =\ \bd{H}\bd{H}\ =\ 0\,.
\label{prop 2}
 \end{align}
\end{subequations}
Using $Q^+\phi=0$ and the general property of the coderivation,
we can show that Eq.~(\ref{recursion 1}) is equivalent to
\begin{equation}
 \bd{H}\bd{L}(e^{\wedge\phi})\ =\ 0\,.
\end{equation}
The solution $\phi_{cl}$ of this equation can be solved for 
$\phi_0$ as
\footnote{The relation (\ref{sol 1}) shows that the map 
$\bd{i}'\equiv(\hat{\bd{I}}-\bd{H}\bd{L}_{int})^{-1}\hat{\bd{P}}$
(deformed inclusion map in the language of HPT) is a cohomomorphism.
This was first shown in Ref.~\cite{Erbin:2020eyc} for the first couple of 
orders in $\bd{L}_{int}$\,, and subsequently proven in Ref.~\cite{Arvanitakis:2020rrk}. 
We also provide an independent proof in Appendix~\ref{app 1}.
}

 \begin{equation}
e^{\wedge\phi_{cl}}\ =\ 
(\hat{\bd{I}}-\bd{H}\bd{L}_{int})^{-1}
\hat{\bd{P}}(e^{\wedge\phi_0})\,,
\label{sol 1}
\end{equation}
and thus
\begin{equation}
 \phi_{cl}\ =\ \pi_1(\hat{\bd{I}}-\bd{H}\bd{L}_{int})^{-1}
\hat{\bd{P}}(e^{\wedge\phi_0})\,.
\label{cl sol}
\end{equation}
The S-matrix generating functional is a functional of $\phi_0$ 
obtained by evaluating the gauge fixed
action (\ref{gf action}) at the classical solution (\ref{cl sol}),
and is given by
\begin{subequations}\label{S gen func}
 \begin{align}
 S[\phi_0]\ \equiv&\ I[\phi_{cl}]
\nonumber\\
=&\ \sum_{n=0}^\infty\frac{1}{(n+2)!}
\Omega(\phi_0, S_{n+1}(\phi_0\,,\cdots\,,\phi_0))\,,
\label{s gen func}
\end{align}
with
\begin{equation}
\bd{S}\ =\
\bd{Q}\hat{\bd{P}}+\bd{S}_{int}\,,\qquad
\bd{S}_{int}\ =\ 
\hat{\bd{P}}\bd{L}_{int}(\hat{\bd{I}}-\bd{H}\bd{L}_{int})^{-1}\hat{\bd{P}}\,,
\label{S matrix}
\end{equation}
\end{subequations}
as is derived in Appendix~\ref{app S matrix}.
This can also be represented by using the multi-linear map\footnote{
Note that the quadratic term in $S[\phi_0]$ vanishes due to
the on-shell property of $\phi_0$\,.
} 
$\langle S|=\sum_{n=0}^\infty\langle S_{(n+3)}|$ with
\begin{subequations}  
\begin{alignat}{3}
\langle S|\ =&\ \langle\Omega|P_0\otimes\pi_1\bd{S}_{int}\,:\quad&\,
&\mathcal{H}_0\otimes \mathcal{SH}_0
\ \longrightarrow\ \mathbb{C}\,,
\label{S generating}\\
\langle S_{n+3}|\ =&\ \langle\Omega|P_0\otimes\pi_1\bd{S}_{int}\pi_{n+2}\,:\quad&\,
&\mathcal{H}_0\otimes(\mathcal{H}_0)^{\wedge (n+2)}\ \longrightarrow\ \mathbb{C}\,.
 \end{alignat}
\end{subequations}
The amplitude for $(n+3)$-string scattering can be obtained as
\begin{equation}
S_{n+3}(\phi_1,\cdots,\phi_{n+3})\
=\ \langle S_{n+3}|(\phi_1\,\otimes\, \phi_2\wedge\cdots\wedge\phi_{n+3})\,,
\end{equation}
where $\phi_i\in\mathcal{H}_0$ are the wave functions (the vertex operators in the
first-quantized formulation) of external string states\,.
We can discuss various $(n+3)$-string scattering amplitudes 
together using $\langle S_{n+3}|$
(or all the scattering amplitudes using $\langle S|$\,).
As an example, the various four-string scattering amplitudes are represented by
\begin{align}
 \langle S_{4}|\ =&\ 
\langle\Omega|P_0\otimes
P_0\pi_1(\bd{L}_3-\bd{L}_2(\id\wedge Q^+\pi_1\bd{L}_2))P_3\pi_3\,.
\end{align}
Taking into account the fact that
the output of the product $\pi_1\bd{L}$ 
satisfies the closed string constraints in Eq.~(\ref{restrict closed})
and has off-shell momentum for inputs with generic momenta,  
we can rearrange it as
\begin{equation}
 \langle S_{4}|\ =\ \langle\omega_s|P_0\otimes P_0\pi_1\bigg(
\bd{l}_3+\bd{l}_2\Big(\id\wedge\left(\frac{b_0^+\mathcal{G}}{L_0^+}\right)\pi_1\bd{l}_2\Big)
\bigg)P_3\pi_3\,,
\end{equation}
from which we can see that the specific amplitude for each process agrees with
that calculated by using the Feynman rules in Ref.~\cite{Kunitomo:2019glq}.

The S-matrix considered here is the total S-matrix in the covariant BRST formalism. 
The physical S-matrix is obtained by projecting it
(or equivalently the external states)
onto the physical subspace 
$\mathcal{H}_Q\subset\mathcal{H}_0$
defined by the (relative) BRST cohomology 
\cite{Kato:1982im,Ohta:1985af,Ito:1985qa,Henneaux:1987ux}:
\begin{equation}
\langle S^{phys}|\ =\ \langle S|\hat{\bd{\mathcal{P}}}\,, 
\end{equation}
with
\begin{equation}
\mathcal{P}\,:\,\mathcal{H}_0\ \longrightarrow\ \mathcal{H}_Q\,,\qquad
\hat{\bd{\mathcal{P}}}\ =\ \sum_{n=0}^\infty\frac{1}{n!}
\underbrace{\mathcal{P}\wedge\cdots\wedge\mathcal{P}}_n\,.
\end{equation}
Decoupling of unphysical states from the physical S-matrix 
is guaranteed by
\begin{align}
 [\bd{Q}\,,\,\bd{S}_{int}]\ =&\ 
\hat{\bd{P}}\Big([\bd{Q}\,,\,\bd{L}_{int}] 
- \bd{L}_{int}(\hat{\bd{I}}-\bd{H}\bd{L}_{int})^{-1}
[\bd{Q}\,,\,\bd{H}\bd{L}_{int}]\Big)
(\hat{\bd{I}}-\bd{H}\bd{L}_{int})^{-1}\hat{\bd{P}}
\nonumber\\
=&\ \hat{\bd{P}}\bd{L}_{int}(\hat{\bd{I}}-\bd{H}\bd{L}_{int})^{-1}
\hat{\bd{P}}\bd{L}_{int}(\hat{\bd{I}}-\bd{H}\bd{L}_{int})^{-1}
\hat{\bd{P}}\ =\ 0\,.
\end{align}
The last equality follows from the fact that the internal states are generically
off-shell.

Here, we have derived the S-matrix generating functional
(\ref{S gen func}) following the physical consideration given 
in Ref.~\cite{Jevicki:1987ax}.
However, it can also be obtained as an (almost) minimal model of the the cyclic 
$L_\infty$ algebra $(\mathcal{H}^{res},\Omega,\bd{L})$ by means of 
HPT \cite{Kajiura:2003ax,Kontsevich:1997vb}. 
This alternative derivation is given in Appendix~\ref{app 1}.
%
%

\subsection{Evaluation of the S-matrix}\label{hetero eval}


Using the multilinear representation (\ref{S generating}),
we can show that all the tree-level amplitudes of the heterotic string
field theory agree with those calculated in the first-quantized formulation.
%

Let us define two maps on the symmetrized tensor algebra $\mathcal{SH}^{res}$\,:
\begin{equation}
 \hat{\bd{i}}'\ =\ (\hat{\bd{I}}-\bd{H}\bd{L}_{int})^{-1}\hat{\bd{P}}\,,
\qquad
 \bd{\Sigma}\ =\ \bd{l}\hat{\bd{i}}'\,.
\end{equation}
Here, as shown in Appendix~\ref{app 1}, 
the first map $\hat{\bd{i}}'$ is a cohomomorphism appeared as a deformed 
chain map in the HPT, 
and thus determined by its component $\pi_1\hat{\bd{i}}'$\,:\,
$\mathcal{SH}^{res}\rightarrow\mathcal{H}^{res}$\,.
Another map $\bd{\Sigma}$ is related with $\bd{S}_{int}$ in (\ref{S matrix})
and $\langle S|$ in (\ref{S generating}) as 
\begin{equation}
\pi_1\bd{S}_{int}\ =\ \mathcal{G}P_0\pi_1\bd{\Sigma}\,,\qquad
\langle S|\ 
=\ \langle\omega_l|\xi_0P_0\otimes P_0\pi_1\bd{\Sigma}\,.
\label{S generating Sigma}
\end{equation}
We can deduce a relation, 
\begin{equation}
 \pi_1\hat{\bd{i}}'\ =\ 
P_0\pi_1-Q^+\mathcal{G}\pi_1\bd{\Sigma}\,,
\end{equation}
between two maps.
Then, from the recursive relation (\ref{recursive l}),
we have the classical Dyson-Schwinger equation
\begin{align}
 \pi_1\bd{\Sigma}\ 
=&\ \sum_{n=0}^\infty \pi_1\bd{B}_{n+2}\left(\frac{1}{(n+2)!}
(\pi_1\hat{\bd{i}}' +
  \Xi\pi_1^1\bd{\Sigma})^{\wedge(n+2)}\right)
\nonumber\\
=&\
\sum_{n=0}^\infty
\pi_1\bd{B}_{n+2}\left(\frac{1}{(n+2)!}
\left(P_0\pi_1
-\Delta\pi_1\bd{\Sigma} \right)^{\wedge(n+2)}\right)\,,
\label{recursive Sigma}
\end{align}
where
\begin{equation}
 \Delta\ =\ Q^+\mathcal{G}-\Xi\pi^1\,.
\end{equation}
We extend the $\bd{\Sigma}$ with two parameters $t$ and $s$ counting 
the picture number and picture number deficit, respectively, so as to satisfy
the generalized Dyson-Schwinger equation
\begin{equation}
  \pi_1\bd{\Sigma}(s,t)\ 
=\
\sum_{n=0}^\infty
\pi_1\bd{B}_{n+2}(s,t)\left(\frac{1}{(n+2)!}
\left(P_0\pi_1
-\Delta(s,t)\pi_1\bd{\Sigma}(s,t) \right)^{\wedge(n+2)}\right)\,,
\label{general recursive Sigma}
\end{equation}
where
\begin{equation}
 \Delta(s,t)\ =\ Q^+\mathcal{G}(s,t)-t\,\Xi\pi^1\,,
\label{ext of prop}
\end{equation}
with $\mathcal{G}(s,t)=\pi^0 + (tX+s)\pi^1$
is determined to satisfy
$\Delta(0,1)=\Delta$ 
and
\begin{equation}
 [\bd{Q},\pi_1\bd{\Sigma}(s,t)]\ 
=\ [\bd{\eta},\pi_1\bd{\Sigma}(s,t)]\ =\ 0\,.
\label{commutativity}
\end{equation}
Then, the derivatives of the extended map $\bd{\Sigma}(s,t)$
satisfy
\begin{equation}
 \pi_1\partial_t\bd{\Sigma}(s,t)\ =\ [\bd{Q},\pi_1\bd{\rho}(s,t)]\,,\qquad
 \pi_1\partial_s\bd{\Sigma}(s,t)\ =\ [\bd{\eta},\pi_1\bd{\rho}(s,t)]\,,
\label{ders of Sigma}
\end{equation}
where concrete form of $\bd{\rho}(s,t)$ is not relevant for the physical S-matrix.
%
The derivations of Eqs.~(\ref{commutativity}) and (\ref{ders of Sigma})
are given in Appendices~\ref{app 2} and \ref{app 3}, respectively.
Combining the two equations (\ref{ders of Sigma}), we can derive the key
relation
\begin{equation}
\pi_1\partial_t\bd{\Sigma}(s,t) - X_0\circ\pi_1\partial_s\bd{\Sigma}(s,t)\
=\ [\bd{Q},[\bd{\eta},\pi_1\bd{T}(s,t)]]\,,
\label{der t der s}
\end{equation}
with $\pi_1\bd{T}(s,t)=\xi_0\circ\pi_1\bd{\rho}(s,t)$. 
Here, for the operator $\mathcal{O}=\xi_0$ or $X_0$\,,
the operation $\mathcal{O}\,\circ$ on 
$\pi_1\bd{D}_n$ is defined by
\begin{equation}
\mathcal{O}\circ\pi_1\bd{D}_n\ =\ \frac{1}{n+1}\Big(
\mathcal{O} D_n+(-1)^{|\mathcal{O}||D|}D_n(\mathcal{O}\pi_1\wedge\id_{n-1})\Big)\,.
\end{equation}

We can find from the definition that the extended S-matrix
$\langle S(s,t)|$ defined by
\begin{equation}
 \langle S(s,t)|\ =\ 
\langle\omega_l|\xi_0P_0\otimes P_0\pi_1\bd{\Sigma}(s,t)
\label{generalized S}
\end{equation}
can be expanded as\footnote{
Note that the lower bound of the summation with respect to $p$ is actually restricted
by the constraint that the cyclic Ramond number ($=$ No. of R-inputs $+$ No. of R-outputs) 
has to be less than the number of external lines ($=$ No. of inputs $+$ No. of outputs).}
\begin{subequations} 
\begin{align}
\langle S(s,t)|\ =&\ \sum_{n=0}^\infty\langle S_{n+3}(s,t)|\,,\\
\langle S_{n+3}(s,t)|\ =&\ \sum_{m=0}^{n+1}s^m
\langle S_{n+3}^{[m]}(t)|\,,\quad 
\langle S_{n+3}^{[m]}(t)|\ =\ \sum_{p=0}^{n-m+1} t^p
\langle S_{n+3}^{(p)}|^{2(n-m-p+1)}\,,
\label{S matrix st}
\end{align}
\end{subequations}
induced by the expansion
\begin{subequations} 
\begin{align}
 \bd{\Sigma}(s,t)\ =&\ \sum_{n=0}^\infty\bd{\Sigma}_{n+2}(s,t)\,,
\label{Sigma expanded}\\
 \bd{\Sigma}_{n+2}(s,t)\ 
=&\ \sum_{m=0}^{n+1} s^m \bd{\Sigma}_{n+2}^{[m]}(t)\,,\quad
\bd{\Sigma}_{n+2}^{[m]}(t)\,,\quad 
=\ \sum_{p=0}^{n-m+1} t^p
\bd{\Sigma}_{n+2}^{(p)}|^{2(n-m-p+1)}\,.
\end{align}
\end{subequations}
The superscripts with $(\cdot)$ and $[\,\cdot\,]$ indicate the picture number
and the picture number deficit, respectively. 
Thus, $\langle S_{n+3}|=\langle S_{n+3}(0,1)|$
is the sum of the amplitudes of several processes as
\begin{equation}
\langle S_{n+3}|\ =\ 
\sum_{p=0}^{n+1} \langle S^{(p)}_{n+3}|^{2(n-p+1)}\,.
\end{equation}
For example, $\langle S_{4}|$ is
the sum of three amplitudes
\begin{align}
\langle S_4|\ =&\ \langle S_4^{(0)}|^4 + \langle S_4^{(1)}|^2
+  \langle S_4^{(2)}|^0\,.
\end{align}
The first and third terms are four-R and four-NS string scattering amplitudes, 
respectively. 
The second term is further the sum of two expressions of the 
two-R-two-NS scattering amplitude,
depending on whether the output (of $\bd{\Sigma}$) is the NS or R state:
\begin{equation}
\langle S_4^{(1)}|^2\ =\ \langle S_4^{(1)}|^2_2 + \langle S_4^{(1)}|^2_0\,.
\end{equation}
Each amplitude can be calculated by solving the Dyson-Schwinger equation
(\ref{recursive Sigma}). We find
\begin{subequations} 
 \begin{align}
 \langle S_4^{(0)}|^4_2\ =&\ \langle\omega_l|\xi_0P_0\otimes P_0\pi_1^1\left(
\bd{B}_3^{(0)}|^4_2-\bd{B}_2^{(0)}|^2_0\left(\id\wedge\frac{b_0^+}{L_0^+}
\pi_1^0\bd{B}_2^{(0)}|^2_2)\right)\right)P_3\pi_3\,,\\
\langle S_4^{(1)}|^2_0\ =&\ \langle\omega_l|\xi_0 P_0\otimes P_0\pi_1^1\bigg(
\bd{B}_3^{(1)}|^2_0
-\bd{B}_2^{(0)}|^2_0\left(\id\wedge\frac{b_0^+}{L_0^+}
\pi_1^0\bd{B}_2^{(1)}|^0_0)\right)
\nonumber\\
&\hspace{4.5cm}
-\bd{B}_2^{(0)}|^2_0\left(\id\wedge\left(\frac{b_0^+X}{L_0^+} -\Xi\right)
\pi_1^0\bd{B}_2^{(0)}|^2_0)\right)\bigg)P_3\pi_3\,,
\label{2NS2R}\\
\langle S_4^{(2)}|^0_0\ =&\ \langle\omega_l|\xi_0P_0\otimes P_0\pi_1^0\left(
\bd{B}_3^{(2)}|^0_0-\bd{B}_2^{(1)}|^0_0\left(\id\wedge\frac{b_0^+}{L_0^+}
\pi_1^0\bd{B}_2^{(1)}|^0_0)\right)\right)P_3\pi_3\,,
\end{align}
which agree with those calculated in Ref.~\cite{Kunitomo:2019glq},
and an alternative expression of the two-NS-two-R amplitude,
\begin{align}
 \langle S_4^{(1)}|^2_2\ =&\ \langle\omega_l|\xi_0P_0\otimes P_0\pi_1^0\bigg(
\bd{B}_3^{(1)}|^2_2
-\bd{B}_2^{(1)}|^0_0\left(\id\wedge\frac{b_0^+}{L_0^+}
\pi_1^0\bd{B}_2^{(0)}|^2_2)\right)
\nonumber\\
&\hspace{4.5cm}
-\bd{B}_2^{(0)}|^2_2\left(\id\wedge\left(\frac{b_0^+X}{L_0^+}
-\Xi\right)
\pi_1^1\bd{B}_2^{(0)}|^2_0)\right)\bigg)P_3\pi_3\,.
\end{align}
\end{subequations}

It is not difficult to prove that the physical S-matrix agrees
with that calculated in the first-quantized formulation.
From (\ref{der t der s}) we find that the extended S-matrix satisfies the relation
\begin{equation}
\partial_t\langle S(s,t)|\ - \partial_s\langle S(s,t)|(X_0)_{cyc}
=\ \langle\omega_l|\xi_0 P_0\otimes P_0\pi_1[\bd{Q},[\bd{\eta},\bd{T}(s,t)]]\,,
\label{key eq}
\end{equation}
where $\langle S(s,t)|(X_0)_{cyc}$ is defined by
\begin{equation}
 \langle S(s,t)|(X_0)_{cyc}\ =\ \sum_{n=0}^\infty\langle S_{n+3}(s,t)|(X_0\otimes\id_{n+2}
+\id\otimes \id_{n+1}\wedge X_0)\,.
\end{equation}
Then, using the expansion in (\ref{S matrix st}), we can find the relation
between components with different picture numbers:
\begin{equation}
 (p+1)\langle S^{(p+1)}_{n+3}|^{2(n-m-p)}\ -\
(m+1)\langle S^{(p)}_{n+3}|^{2(n-m-p)}(X_0)_{cyc}\
=\ 
\langle\omega_l|\xi_0 P_0\otimes P_0\pi_1[\bd{Q},[\bd{\eta},\ast\,]]\,.
\label{decent eq}
\end{equation}
The explicit form of the right hand side is not important
since it does not contribute to the physical S-matrix thanks to the fact that
$\bd{Q}\hat{\bd{\mathcal{P}}}=
\bd{\eta}\hat{\bd{\mathcal{P}}}=0$\,.
By using this relation repeatedly, we can find the physical amplitude
can eventually be written as
\begin{equation}
 \langle S^{phys}_{n+3}|\ 
=\ \sum_{p=0}^{n+1}\langle S^{(0)}_{n+3}|^{2(n-p+1)}((X_0)_{cyc})^p
\hat{\bd{\mathcal{P}}}\,.
\end{equation}
In this final form, the PCOs in an amplitude are acting on the external states
in a way that is common to all the Feynman diagrams, so
each amplitude is written as integral of a smooth function (section) over the whole moduli space
including external states accompanied by PCOs. 
If we further note that the differences in PCOs we use and in
the states they act can be written in the form of $[Q,[\eta,\ *\ ]]$\,,\footnote{
For example, if we map the worldsheet to the complex $z$ plane so that
the string 1 and 2 are on the points $Z_1$ and $Z_2$, respectively,
the differences can be written as
\begin{align}
 (X_0)_1-(X_0)_2\ =&\ [Q\,,\,[\eta\,,\,
\xi_0\oint_{Z_1}\frac{dz_1}{2\pi iz_1}
\oint_{Z_2}\frac{dz_2}{2\pi iz_2}
\int^{z_1}_{z_2}dw\partial\xi(w)]]\,,\nonumber\\
(X_0)_1-X(Z_1)\ =&\ [Q\,,\,[\eta\,,\,\xi_0\oint_{Z_1}\frac{dz_1}{2\pi i z_1}
\int^{z_1}_{Z_1}dz\partial\xi(z)]]\nonumber\,,
\end{align}
where $(X_0)_1$ and $(X_0)_2$ are $X_0$s acting on the string 1 and 2, respectively.
}
we can replace $X_0$s with the local ones $X(z)$s 
so that all the external NS and R states
have the picture number either $-1$ or $0$ and $-1/2$ or $-3/2$\,, respectively.
Then, 
we can conclude that the heterotic string field theory reproduces the tree-level 
S-matrix calculated in the covariant first-quantized formulation \cite{Friedan:1985ge}.

\section{Extension to the type II superstring field theory}\label{type II}

In the previous section, we show that the tree-level physical S-matrix of the heterotic
string field theory agrees with that calculated in the conventional first-quantized 
formulation. This proof is easily extended to the type II superstring field theory
since it was constructed by repeating the prescription
developed for the heterotic string field theory \cite{Kunitomo:2019kwk}.

\subsection{Type II superstring field theory with $L_\infty$ structure}\label{type II summ}

The type II superstring field $\Phi$ is Grassmann even and has four components
with ghost number 2:
\begin{equation}
 \Phi\ =\ \Phi_{\NSNS} + \Phi_{\RNS}  + \Phi_{\NSR} + \Phi_{\RR} \in\ 
\mathcal{H}^{res}=\mathcal{H}_{\NSNS} + \mathcal{H}_{\RNS}^{res}
+ + \mathcal{H}_{\NSR}^{res} + \mathcal{H}_{\RR}^{res}\,.
\end{equation}
The NS-NS component $\Phi_{\NSNS}$\,, R-NS component $\Phi_{\RNS}$\,,
NS-R component $\Phi_{\NSR}$\,, and R-R component $\Phi_{\RR}$ 
have picture numbers $(-1,-1)$\,, $(-1/2,-1)$\,, $(-1,-1/2)$\,, and $(-1/2,-1/2)$\,,
respectively. The Hilbert space $\mathcal{H}^{res}$ is restricted by
the closed string constraints (\ref{restrict closed})
and an extra condition (\ref{restricted}) with
\begin{alignat}{6}
 \mathcal{G}\ =&\ \mathscr{X}\overline{\mathscr{X}}\,, \qquad&
\mathscr{X}\ =&\ \pi^0+X\pi^1\,, \qquad&
\overline{\mathscr{X}}\ =&\ \overline{\pi}^0+\overline{X}\overline{\pi}^1\,,\\
 \mathcal{G}^{-1}\ =&\ \mathscr{Y}\overline{\mathscr{Y}}\,, \qquad&
\mathscr{Y}\ =&\ \pi^0+\mathcal{Y}\pi^1\,, \qquad&
\overline{\mathscr{Y}}\ =&\ \overline{\pi}^0+\overline{\mathcal{Y}}\overline{\pi}^1\,,
\end{alignat}
where $\pi^0$ and $\pi^1$ ($\overline{\pi}^0$ and $\overline{\pi}^1$) 
are the projection operators onto the NS and R states
in the holomorphic (anti-holomorphic) sectors, respectively. 
We take the non-local operators
\begin{equation}
 \mathcal{Y}\ =\ -2\frac{G}{L_0^+}\delta(\gamma_0)\,,\qquad
 \overline{\mathcal{Y}}\ -2\frac{\overline{G}}{L_0^+}\delta(\overline{\gamma}_0)\,,
\end{equation}
satisfying the same relations as those satisfied by $Y$ and $\overline{Y}$\,,
\begin{equation}
  X\mathcal{Y}X\ =\ X\,,\qquad \mathcal{Y}X\mathcal{Y}\ =\ \mathcal{Y}\,,\qquad 
\overline{X}\,\overline{\mathcal{Y}}\,\overline{X}\ =\ \overline{X}\,,\qquad
\overline{\mathcal{Y}}\,\overline{X}\,\overline{\mathcal{Y}}\ =\ \overline{\mathcal{Y}}\,,
\end{equation}
as inverse picture changing operators, 
which is necessary to consistently impose all the constraints.\footnote{
In Ref.~\cite{Kunitomo:2019kwk}, we took these non-local inverse changing
operators only for the R-R sector, which is enough for consistency.
The result is the same in both cases if we expand the string field in the ghost
zero-modes as in Eq.~(\ref{zero mode form type II}).} 
The constraint (\ref{restricted}) can now be split into two conditions
\begin{equation}
\mathscr{X}\mathscr{Y}\Phi\ =\ \overline{\mathscr{X}}\,\overline{\mathscr{Y}}\Phi=\Phi\,,
\end{equation}
which we use later.
%
Any sate $\mathcal{B}\in\mathcal{H}^{res}$ can be expanded in the ghost
zero-mode as
\begin{align}
 \mathcal{B} =&\ \mathcal{B}_{\NSNS} + \mathcal{B}_{\RNS}
+ \mathcal{B}_{\NSR} + \mathcal{B}_{\RR}
\nonumber\\
=&\
\left(b_{NS}-c_0^+B_{\NSNS}\right) 
+ \left(b_{\RNS} - \frac{1}{2}(\gamma_0+2c_0^+G)B_{\RNS}\right)
\nonumber\\
&\
+ \left(b_{\NSR} - \frac{1}{2}(\overline{\gamma}_0+2c_0^+\overline{G})B_{\NSR}\right)
+ \left(b_{\RR} 
- \frac{1}{2}(\gamma_0\overline{G}-\overline{\gamma}_0G+2c_0^+G\overline{G})B_{\RR}\right),
\label{zero mode form type II}
\end{align}
so we denote the string field $\Phi\in\mathcal{H}^{res}$ as
\begin{align}
 \Phi =&\ \Phi_{\NSNS} + \Phi_{\RNS}
+ \Phi_{\NSR} + \Phi_{\RR}
\nonumber\\
=&\
\left(\phi_{NS}-c_0^+\psi_{\NSNS}\right) 
+ \left(\phi_{\RNS} - \frac{1}{2}(\gamma_0+2c_0^+G)\psi_{\RNS}\right)
\nonumber\\
&\
+ \left(\phi_{\NSR} - \frac{1}{2}(\overline{\gamma}_0+2c_0^+\overline{G})\psi_{\NSR}\right)
+ \left(\phi_{\RR} 
- \frac{1}{2}(\gamma_0\overline{G}-\overline{\gamma}_0G+2c_0^+G\overline{G})\psi_{\RR}\right).
\label{zero mode form type II field}
\end{align}
The on-shell subspace $\mathcal{H}_0\subset\mathcal{H}^{res}$ 
is defined by
\begin{align}
 \mathcal{H}_0\ =\ 
\{\Phi\in\mathcal{H}^{res}\,|\,L_0^+\Phi_{\NSNS}=G_0\Phi_{\RNS}
=\overline{G}_0\Phi_{\NSR}=G_0\Phi_{\RR}=\overline{G}_0\Phi_{\RR}=0\}\,.
\end{align}
The natural symplectic forms 
$\Omega$\,, $\omega_s$\,, and $\omega_l$ 
in the $\mathcal{H}^{res}=\mathcal{H}^{res}_L\otimes\mathcal{H}^{res}_R$\,,
$\mathcal{H}_s=(\mathcal{H}_s)_L\otimes(\mathcal{H}_s)_R$\,,
and
$\mathcal{H}_l=(\mathcal{H}_l)_L\otimes(\mathcal{H}_l)_R$\,,
respectively, are defined similarly to those in Eqs.(\ref{symplectic}), 
and are related for $\Phi_1,\Phi_2\in\mathcal{H}^{res}$ as
\begin{equation}
 \Omega(\Phi_1\,,\Phi_2)\ =\ \omega_s(\Phi_1\,,\mathcal{G}^{-1}\Phi_2)\
=\ \omega_l(\xi_0\overline{\xi}_0\Phi_1,\mathcal{G}^{-1}\Phi_2)\,.
\end{equation}
It is also useful to introduce 
$\Omega_m$ in the \textit{restricted medium} Hilbert space
$\mathcal{H}_m^{res}=(\mathcal{H}_l)_L\otimes\mathcal{H}^{res}_R$
 and $\omega_m$ in the \textit{medium} Hilbert space
$\mathcal{H}_m=(\mathcal{H}_l)_L\otimes(\mathcal{H}_s)_R$\,, 
and their counterparts the holomorphic and anti-holomorphic sectors exchanged
$\Omega_{\overline{m}}$ in
$\mathcal{H}_{\overline{m}}^{res}=\mathcal{H}^{res}_L\otimes(\mathcal{H}_l)_R$
and $\omega_{\overline{m}}$ in $\mathcal{H}_{\overline{m}}=(\mathcal{H}_s)_L\otimes(\mathcal{H}_l)_R$\,,
which are defined in an obvious way.
They are related with $\omega_l$ as
\begin{alignat}{4}
 \Omega_m(\Phi_1\,,\Phi_2)\ =&\ 
\omega_m(\Phi_1\,,\overline{\mathscr{Y}}\Phi_2)\ =&\
\omega_l(\overline{\xi_0}\Phi_1\,,\overline{\mathscr{Y}}\Phi_2)\,,\qquad
\Phi_1,\Phi_2\in&\mathcal{H}_m^{res}\,,\\
 \Omega_{\overline{m}}(\Phi_1\,,\Phi_2)\ =&\ 
\omega_{\overline{m}}(\Phi_1\,,\mathscr{Y}\Phi_2)\ =&\
\omega_l(\xi_0\Phi_1\,,\mathscr{Y}\Phi_2)\,,\qquad
\Phi_1,\Phi_2\in&\mathcal{H}_{\overline{m}}^{res}\,,
\end{alignat}
and with $\Omega$ as
\begin{align}
\Omega_m(\xi_0\Phi_1\,,\mathscr{Y}\Phi_2)\ =\
\Omega_{\overline{m}}(\overline{\xi_0}\Phi_1\,,\overline{\mathscr{Y}}\Phi_2)\
=\ 
 \Omega(\Phi_1\,,\Phi_2)\,,\qquad
\Phi_1,\Phi_2\in\mathcal{H}^{res}\,.
\end{align}
The action and the gauge transformation have the same form as
those of the heterotic string field theory,
\begin{subequations} 
\begin{align}
 I[\Phi]\ =&\ \sum_{n=0}^\infty\frac{1}{(n+2)!}\,
\Omega(\Phi\,,L_{n+1}(\underbrace{\Phi\,,\cdots\,,\Phi}_{n+1}))\,,
\label{small action II}\\
 \delta\Phi\ =&\ \sum_{n=0}^\infty\frac{1}{n!}\, 
L_{n+1}(\underbrace{\Phi\,,\cdots\,,\Phi}_n\,,\Lambda)\,,
\label{small gauge tf II}
\end{align} 
\end{subequations}
except that 
the symplectic form $\Omega$ and
the string products $L_{n+1}$ 
are now those for the type II superstring field theory. 

The string products are constructed by repeating 
twice the prescription developed for the heterotic string products.
We first apply it to the holomorphic
sector and obtain the \textit{heterotic product}
$\bd{L}_H=\bd{Q}+\bd{L}_H^{\,int}$ with
\begin{subequations} 
 \begin{align}
 \bd{L}_H\ =&\ \hat{\bd{F}}^{-1}(\bd{Q}+\pi^0\bd{B})\hat{\bd{F}}\,,\\
n \pi_1\bd{L}_H^{int}\ =&\ \mathscr{X}\pi_1\bd{l}_H\,,\qquad
\pi_1\bd{l}_H\ =\ \pi_1\bd{B}\hat{\bd{F}}\,.
\end{align}
\end{subequations}
All the quantities and relations in this first step
have the same form as those of the heterotic string field theory in appearance.
Next, in the second step, we repeat the prescription for the anti-holomorphic
sector and construct the type II superstring product as
\begin{subequations} 
 \begin{align}
\bd{L}\ =&\ \hat{\overline{\bd{F}}}^{-1}(\bd{Q}+\overline{\pi}^0\overline{\bd{B}})\hat{\overline{\bd{F}}}\,,\\
 \pi_1\bd{L}_{int}\ =&\ \overline{\mathscr{X}}\pi_1\overline{\bd{l}}\,,\qquad 
\pi_1\overline{\bd{l}}\ =\ \pi_1\overline{\bd{B}}\,\hat{\overline{\bd{F}}}\,.
\label{type II product}
\end{align} 
\end{subequations}
The product $\overline{\bd{B}}$ is obtained from the generating function
\begin{equation}
 \overline{\bd{B}}(s,t)\ =\ \sum_{\overline{m},\overline{n},\overline{r}=0}^\infty
s^{\overline{m}}t^{\overline{n}}\,\overline{\bd{B}}^{(\overline{n})}_{\overline{m}+\overline{n}+\overline{r}+1}|^{2\overline{r}}\
\equiv \sum_{\overline{n}=0}^\infty t^{\overline{n}}\,\overline{\bd{B}}^{(\overline{n})}(s)\,,
\end{equation}
with the gauge product
\begin{equation}
 \overline{\bd{\lambda}}(s,t)\ =\ \sum_{\overline{m},\overline{n},\overline{r}=0}^\infty 
s^{\overline{m}} t^{\overline{n}}\, 
\overline{\bd{\lambda}}^{(\overline{n}+1)}_{\overline{m}+\overline{n}+\overline{r}+2}|^{2\overline{r}}\,,
\end{equation}
by solving the differential equation
\begin{subequations}\label{diff eq type II} 
 \begin{align}
 \partial_t\overline{\bd{B}}(s,t)\ =&\ [\bd{Q},\overline{\bd{\lambda}}(s,t)]
+ [\overline{\bd{B}}(s,t),\overline{\bd{\lambda}}(s,t)]^{\overline{1}}
  + s\, [\overline{\bd{B}}(s,t),\overline{\bd{\lambda}}(s,t)]^{\overline{2}}\,,\\
\partial_s\overline{\bd{B}}(s,t)\ =&\ [\overline{\bd{\eta}},\overline{\bd{\lambda}}(s,t)]
- t\, [\overline{\bd{B}}(s,t),\overline{\bd{\lambda}}(s,t)]^{\overline{2}}\,,
\end{align}
\end{subequations}
with the initial condition 
\begin{equation}
\overline{\bd{B}}^{(\overline{0})}(s)\ =\ \bd{L}_H^{\,int}(s)\ =\ 
\sum_{\overline{m},\overline{r}=0}^\infty s^{\overline{m}}(\bd{L}_H)_{\overline{m}+\overline{r}+1}|^{2\overline{r}}\,.
\end{equation}
The cohomomorphism $\hat{\overline{\bd{F}}}$ is given by
\begin{equation}
 \hat{\overline{\bd{F}}}^{-1}\ =\ \pi_1\id - \overline{\Xi}\overline{\pi}^1_1\overline{\bd{B}}\,,
\end{equation}
where $\overline{\pi}_1^1=\pi_1\overline{\pi}^1$\,.
Here, the picture and cyclic Ramond numbers are those of the anti-holomorphic sector,
which are made clear by letters with bar. The commutators $[\cdot\,,\cdot]^{\overline{1},\overline{2}}$ 
are the analogues of $[\cdot\,,\cdot]^{1,2}$ 
projected by the cyclic Ramond number of the anti-holomorphic sector.
It would be obvious that $\overline{\bd{B}}(s,t)$ satisfies the similar relations to
Eqs.~(\ref{extended L infty}) as
\begin{subequations}\label{extended L infty antiholo} 
\begin{align}
&\ [\bd{Q},\overline{\bd{B}}(s,t)] + \frac{1}{2}\,[\overline{\bd{B}}(s,t),\overline{\bd{B}}(s,t)]^{\overline{1}}
+ \frac{s}{2}\,[\overline{\bd{B}}(s,t),\overline{\bd{B}}(s,t)]^{\overline{2}}\ =\ 0\,,\\
&\ [\overline{\bd{\eta}},\overline{\bd{B}}(s,t)] - \frac{t}{2}\,[\overline{\bd{B}}(s,t),\overline{\bd{B}}(s,t)]^2\ =\ 0\,,
\end{align}
\end{subequations}
which reduce to the $L_\infty$ relation 
of the coderivation $\bd{Q}-\overline{\bd{\eta}}+\overline{\bd{B}}$ at $(s,t)=(0,1)$\,.
If you note that the initial heterotic product has the form of 
$\bd{L}_H^{\,int}=\mathscr{X}\bd{l}_H$ and look carefully 
the differential equation (\ref{diff eq type II}), it is found that the final form
of $\overline{\bd{B}}$ has the form of
\begin{equation}
 \pi_1\overline{\bd{B}}\ =\ \mathscr{X}\pi_1\overline{\bd{b}}\,,
\end{equation}
and thus $\pi_1\bd{L}_{int}$ is written as the form in which
it is manifest that the products is closed in $\mathcal{H}^{res}$\,:
\begin{equation}
  \pi_1\bd{L}_{int}\ =\ \mathcal{G}\pi_1\bd{l}\,,\qquad
\pi_1\bd{l}\ =\ \pi_1\overline{\bd{b}}\,\hat{\overline{\bd{F}}}\,.
\label{type II product 2}
\end{equation}

\subsection{S-matrix generating function and its evaluation}\label{type II eval}

Since the equation of motion has the same form as that of the heterotic string
field theory, the S-matrix generating function can also be written in the same form
as (\ref{S generating}) with
\begin{equation}
 \bd{S}_{int}\ =\ 
\hat{\bd{P}}\bd{L}_{int}(\hat{\bd{I}}-\bd{H}\bd{L}_{int})^{-1}\hat{\bd{P}}\,,
\end{equation}
where $\bd{L}_{int}$ is the type II superstring product given by 
(\ref{type II product 2}).
By repeating the proof for the heterotic string amplitudes,
we can again show that the physical amplitudes of type II superstring are obtained as 
integrals of the smooth functions (sections) over the whole moduli space with
appropriate picture changed external states.

Let us briefly summarize the procedure for the type II superstring.
If we define the maps
\begin{equation}
 \hat{\bd{i}}'\ =\ (\hat{\bd{I}}-\bd{H}\bd{L}_{int})^{-1}\hat{\bd{P}}\,,\qquad
\pi_1\overline{\bd{\Sigma}}\ =\ \pi_1\overline{\bd{l}}\hat{\bd{i}}'\,,
\end{equation}
the S-matrix can be represented by $\overline{\bd{\Sigma}}$
similar to that of the heterotic string field theory as
\begin{align}
\pi_1\bd{S}_{int}\ =&\ \overline{\mathscr{X}}P_0\pi_1\overline{\bd{\Sigma}}\,,\\
\langle S|\ =&\ \langle\Omega|P_0\otimes\pi_1\bd{S}_{int}\ =\ 
\langle\Omega_{\overline{m}}|\overline{\xi}_0P_0\otimes P_0\pi_1\overline{\bd{\Sigma}}\,.
\end{align}
The $\overline{\bd{\Sigma}}$
is recursively determined by the equation
\begin{equation}
 \pi_1\overline{\Sigma}\ =\ 
\sum_{n=0}^\infty\pi_1\overline{\bd{B}}_{n+2}\bigg(
\frac{1}{(n+2)!}(P_0\pi_1-\overline{\Delta}\pi_1\bd{\Sigma})^{\wedge(n+2)}
\bigg)\,,
\end{equation}
with 
\begin{equation}
 \overline{\Delta}=Q^+\overline{\mathscr{X}}-\overline{\Xi}\overline{\pi}^1\,.
\end{equation}
If we define two parameter extension $\overline{\bd{\Sigma}}(s,t)$ 
using $\overline{\bd{B}}(s,t)$ and
\begin{equation}
 \overline{\Delta}(s,t)\ =\ Q^+\overline{\mathscr{X}}(s,t)-t\,\overline{\Xi}\overline{\pi}^1\,,\qquad
\overline{\mathscr{X}}(s,t)\ =\ \overline{\pi}^0+(t\overline{X}+s)\overline{\pi}^1\,,
\end{equation}
it is easy to see that the relations
\begin{equation}
 \partial_t\overline{\bd{\Sigma}}(s,t)\ =\ [\bd{Q},\overline{\bd{\rho}}(s,t)]\,,\qquad
 \partial_s\overline{\bd{\Sigma}}(s,t)\ =\ [\overline{\bd{\eta}},\overline{\bd{\rho}}(s,t)]\,,
\label{der Sigma type II}
\end{equation}
hold similar to the case of the heterotic string field theory.
The explicit form of $\overline{\bd{\rho}}(s,t)$ is not relevant for the physical
S-matrix, but obtained similarly to that of the heterotic string given in
Appendix~\ref{app 3}.
The relations (\ref{der Sigma type II}) provides the key equation
\begin{equation}
 \partial_t\overline{\bd{\Sigma}}(s,t)
-\overline{X}_0\circ\partial_s\overline{\bd{\Sigma}}(s,t)\
=\ [\bd{Q},[\overline{\bd{\eta}},\overline{\bd{T}}(s,t)]]\,,
\end{equation}
with $\overline{\bd{T}}(s,t)=\overline{\xi}_0\circ\overline{\bd{\rho}}(s,t)$\,.
Then, the extended S-matrix 
\begin{align}
 \langle S_{n+3}(s,t)|\ =&\ \sum_{\overline{m}=0}^{n+1}s^{\overline{m}}
\langle S_{n+3}^{[\overline{m}]}(t)|\,,\quad 
\langle S_{n+3}^{[\overline{m}]}(t)|\ =\ \sum_{\overline{p}=0}^{n-\overline{m}+1} 
t^{\overline{p}}\langle S_{n+3}^{(\overline{p})}|^{2(n-\overline{m}-\overline{p}+1)}\,,
\label{S matrix st type II}
\end{align}
induced by the expansion
\begin{equation}
 \overline{\bd{\Sigma}}_{n+2}(s,t)\ 
=\ \sum_{\overline{m}=0}^{n+1} s^{\overline{m}}\, \overline{\bd{\Sigma}}_{n+2}^{[\overline{m}]}(t)\,,\quad
\overline{\bd{\Sigma}}_{n+2}^{[\overline{m}]}(t)\,,\quad 
=\ \sum_{\overline{p}=0}^{n-\overline{m}+1} t^{\overline{p}}\,
\overline{\bd{\Sigma}}_{n+2}^{(\overline{p})}|^{2(n-\overline{m}-\overline{p}+1)}\,,
\end{equation}
satisfies the relation
\begin{equation}
\partial_t\langle S(s,t)|\ - \partial_s\langle S(s,t)|\overline{X}_0
=\ \langle\Omega_m|\overline{\xi}_0 P_0\otimes P_0\pi_1[\bd{Q},
[\overline{\bd{\eta}},\overline{\bd{T}}(s,t)]]\,.
\label{key eq type II}
\end{equation}
Hence, for the physical S-matrix, 
we have
\begin{equation}
 \langle S^{phys}_{n+3}|\ 
=\ \sum_{\overline{p}=0}^{n+1}\langle S_{n+3}^{(\overline{0})}|^{2(n-\overline{p}+1)}
(\overline{X}_0)^{\overline{p}}\hat{\bd{\mathcal{P}}}\,,
\label{1st step result}
\end{equation}
in a similar way to the case of the heterotic string field theory.
Here, $\hat{\bd{\mathcal{P}}}$ is the projection operator
onto the physical Hilbert space satisfying $\bd{Q}\hat{\bd{\mathcal{P}}}=
\bd{\eta}\hat{\bd{\mathcal{P}}}=\overline{\bd{\eta}}\hat{\bd{\mathcal{P}}}=0$\,.
The amplitudes with no anti-holomorphic picture number
$\langle S_{n+3}^{(\overline{0})}|$
is nothing but those of the heterotic string 
field theory,
\begin{equation}
\langle S_{n+3}^{(\overline{0})}|\ =\
 \langle (S_H)_{n+3}|\,,
\end{equation}
by construction. 
It is embedded in the type II superstring Hilbert space as
\begin{equation}
 \langle S_H|\ 
=\ \langle\Omega_m|\xi_0P_0\otimes \mathscr{Y}\pi_1\bd{S}_H\
=\ \langle\Omega_m|\xi_0P_0\otimes P_0\pi_1\bd{\Sigma}_H\,,
\end{equation}
with
\begin{equation}
 \bd{S}_H\ =\ \hat{\bd{P}}\bd{L}_H^{int}(\hat{\bd{I}}-\bd{H}\bd{L}_H^{int})^{-1}\hat{\bd{P}}\,.
\end{equation}
So, now it is easy to see that 
\begin{equation}
 \langle (S_H)_{n+3}|\ =\ \sum_{p=0}^{n+1}\langle S^{(0)}_{n+3}|^{2(r-p+1)}(X_0)^p\,,
\end{equation}
without explicitly repeating the procedure again. Substituting this into the first
step result (\ref{1st step result}), the physical S-matrix of the type II 
superstring field theory can be written as 
\begin{align}
 \langle S_{n+3}^{phys}|\ =\
\sum_{p=0}^{n+1}\sum_{\overline{p}=0}^{n+1}
\langle S_{n+3}^{(0,\overline{0})}|^{\left(2(n-p+1),2(n-\overline{p}+1)\right)}
(X_0)^p(\overline{X}_0)^{\overline{p}}
\hat{\bd{\mathcal{P}}}\,.
\end{align}
From the similar consideration to that in the heterotic string field theory,
this agrees with the tree-level
physical $(n+3)$-string scattering amplitudes 
obtained in the covariant first-quantized formulation.

\section{Extension to the open superstring field theory}\label{open}

The proof given for closed superstring theories in the previous sections 
can also be applied to the case of the open superstring field theory.
In this section, we first extend our construction method
to the open superstring field theory, and then,
prove that it reproduces the well-known tree-level S-matrix.

\subsection{Open superstring field theory with cyclic $A_\infty$ structure}\label{open summ}

The open superstring field $\Phi$ is Grassmann odd and has two components with ghost
number 1:
\begin{equation}
 \Phi\ =\ \Phi_{NS} + \Phi_R\ \in \mathcal{H}^{res}\ =\ \mathcal{H}_{NS}+\mathcal{H}^{res}_R\,.
\end{equation}
The NS component $\Phi_{NS}$ and R component $\Phi_R$ have picture
numbers $-1$ and $-1/2$\,, respectively. The Hilbert space $\mathcal{H}^{res}$ is
restricted by a constraint (\ref{restricted}) with
\begin{subequations} 
\begin{alignat}{3}
  \mathcal{G}\ =&\ \pi^0 + X \pi^1\,,\qquad&
 X\ =&\ -\delta(\beta_0)G_0 + \delta'(\beta_0)b_0\,,\\
 \mathcal{G}^{-1}\ =&\ \pi^0 + Y \pi^1\,,\qquad&
Y\ =&\ -c_0\delta'(\gamma_0)\,.
\end{alignat}
\end{subequations}
The string field $\Phi\in\mathcal{H}^{res}$ is expanded in the ghost
zero-modes as
\begin{align}
 \Phi\ =\ (\phi_{NS}-c_0\psi_{NS}) + \left(\phi_{R}-\frac{1}{2}(\gamma_0+c_0G)\psi_R\right)\,.
\end{align}
The on-shell subspace $\mathcal{H}_0\subset\mathcal{H}^{res}$ and the projector $P_0$
are introduced as
\begin{equation}
P_0\,:\, \mathcal{H}^{res}\longrightarrow 
\mathcal{H}_0\ =\ \{\Phi\in{\mathcal{H}^{res}\mid L_0\Phi_{NS}\ =\ G_0\Phi_R\ =\ 0\,}\}\,.
\label{open onshell}
\end{equation}
The symplectic forms are defined by
\begin{subequations} \label{symplectic o}
 \begin{alignat}{4}
 \omega_l(\Phi_1,\Phi_2)\ =&\ (-1)^{\textrm{deg}(\Phi_1)}{}_l\langle \Phi_1| \Phi_2\rangle_l\,,
\qquad& \Phi_1,\Phi_2\in&\ \mathcal{H}_l\,,\label{symp l o}\\
 \omega_s(\Phi_1,\Phi_2)\ =&\ (-1)^{\textrm{deg}(\Phi_1)}{}_s\langle \Phi_1|\Phi_2\rangle_s\,,
\qquad& \Phi_1,\Phi_2\in&\ \mathcal{H}_s\,,\label{symp s o}\\
 \Omega(\Phi_1,\Phi_2)\ =&\ 
(-1)^{\textrm{deg}(\Phi_1)}{}_s\langle \Phi_1|\mathcal{G}^{-1}|\Phi_2\rangle_s\,,\qquad& 
\Phi_1,\Phi_2\in&\ \mathcal{H}^{res}\,,\label{symp res o}
\end{alignat}
\end{subequations}
using the BPZ inner product.
The degree of generic states $\Phi_{1,2}$ are defined by 
$\textrm{deg}(\Phi_{1,2})=(|\Phi_{1,2}|+1)\ \textrm{mod}\ 2$\,, 
and thus, in particular, the degree of the open superstring field $\Phi$ is even.

The action and gauge transformation 
are now given by
\begin{equation}
 I[\Phi]\ =\ \sum_{n=0}^\infty \frac{1}{n+2}
\Omega(\Phi,M_{n+1}(\underbrace{\Phi,\cdots,\Phi}_{n+1}))\,,
\end{equation}
and
\begin{equation}
 \delta\Phi\ =\ \sum_{n=0}^\infty\sum_{k=0}^n
M_{n+1}(\underbrace{\Phi,\cdots,\Phi}_{n-k},\Lambda,\underbrace{\Phi,\cdots,\Phi}_{k})\,,
\end{equation}
by using the non-commutative open-superstring products, $M_1=Q$  and $M_{n+2}\ (n\ge0)$\,, 
which satisfy the $A_\infty$ relations
\begin{align}
\begin{split}
&\ \sum_{m=0}^n\sum_{j=0}^{n-m}
(-1)^{\epsilon(1,j)}\ 
\\
&\hspace{10mm} 
\times\ M_{n-m+1}(\Phi_1\,,\cdots,\Phi_j\,,M_{m+1}(\Phi_{j+1}\,,\cdots,\Phi_{j+m+1})\,,
\Phi_{j+m+2}\,,\cdots,\Phi_{n+1})\ =\ 0\,,
\end{split}
\end{align}
and the cyclicity condition
\begin{equation}
\Omega(M_n(\Phi_1\,,\cdots\,,\Phi_n),\Phi_{n+1})\ =\ 
(-1)^{\textrm{deg}(\Phi_1)\epsilon(2,n+1)}
\Omega(M_n(\Phi_2\,,\cdots\,,\Phi_{n+1}),\Phi_1)\,,
\end{equation}
where $\epsilon(i,j)=\sum_{k=i}^{j}\textrm{deg}(\Phi_k)$\,.

Our construction method of the string products of the heterotic string field theory 
is also applicable to those of the open superstring field theory.
It is achieved by simply replacing 
the coderivation $\bd{B}(s,t)$ acting on the symmetrized tensor algebra
$\mathcal{SH}_l$ with 
the coderivation
\begin{equation}
 \bd{A}(s,t)\ =\ \sum_{n,m,r=0}^\infty s^m t^n \bd{A}^{(n)}_{m+n+r+1}|^{2r}\,,
\end{equation}
acting on the tensor algebra 
$\mathcal{TH}^{res}=\oplus_{n=0}^\infty(\mathcal{H}^{res})^{\otimes n}$\,.
If $\bd{A}(s,t)$ satisfies the differential equations
\begin{subequations}\label{diff eq open} 
 \begin{align}
 \partial_t\bd{A}(s,t)\ =&\ [\bd{Q},\bd{\mu}(s,t)]
+ [\bd{A}(s,t),\bd{\mu}(s,t)]^1
  + s\, [\bd{A}(s,t),\bd{\mu}(s,t)]^2\,,\\
\partial_s\bd{A}(s,t)\ =&\ [\bd{\eta},\bd{\mu}(s,t)]
- t\, [\bd{A}(s,t),\bd{\mu}(s,t)]^2\,,
\end{align}
\end{subequations}
with the (even) coderivation of the open superstring gauge products 
\begin{equation}
 \bd{\mu}(s,t)\ =\ \sum_{m,n,r=0}^\infty s^m t^n \bd{\mu}^{(n+1)}_{m+n+r+2}|^{2r}\,,
\end{equation}
it satisfies the similar relations to Eqs.~(\ref{extended L infty}),
\begin{subequations}\label{extended A infty}
\begin{align}
&\ [\bd{Q},\bd{A}(s,t)] + \frac{1}{2}\,[\bd{A}(s,t),\bd{A}(s,t)]^1
+ \frac{s}{2}\,[\bd{A}(s,t),\bd{A}(s,t)]^2\ =\ 0\,,\\
&\ [\bd{\eta},\bd{A}(s,t)] - \frac{t}{2}\,[\bd{A}(s,t),\bd{A}(s,t)]^2\ =\ 0\,.
\end{align}
\end{subequations}
At $(s,t)=(0,1)$\,, they reduce to the $A_\infty$ relations 
of the coderivation $\bd{Q}-\bd{\eta}+\bd{A}(0,1)$\,. 
All the open superstring products, and simultaneously the gauge products,
are determined in the same way as those of the heterotic string field theory by
solving the differential equations in (\ref{diff eq open})
starting from the initial condition 
\begin{equation}
\bd{A}(s,0)\ =\ \bd{M}_B^{(0)}(s)\
=\ \sum_{m,r=0}^\infty s^m \bd{M}^{(0)}_{m+r+1}|^{2r}\,.
\end{equation}
Here, $\bd{M}^{(0)}_{n+2}$
are the open string products without PCO-insertions,
which we call the open bosonic string products. 
The cyclic $A_\infty$ algebra $\bd{M}=\bd{Q}+\bd{M}_{int}$ is constructed from
$\bd{A}=\bd{A}(0,1)$ as
\begin{subequations}\label{A infty}  
\begin{align}
\bd{M}\ =&\ \hat{\bd{F}}^{-1}(\bd{Q}+\pi^0\bd{A})\hat{\bd{F}}\,,\\
\pi_1\bd{M}_{int}\ =&\ \mathcal{G}\bd{a}\,,\qquad
\pi_1\bd{a}\ =\ \pi_1\bd{A}\hat{\bd{F}}\,,
\end{align}  
\end{subequations}
using the cohomomorphism 
\begin{equation}
\pi_1\hat{\bd{F}}^{-1}\ =\ \pi_1\id - \Xi\pi_1^1\bd{A}\,.
\label{open F} 
\end{equation}
This construction is an extension 
of that in Ref.~\cite{Erler:2013xta} proposed for the products in the NS sector.
If we restrict our construction to the NS sector,
the cohomomorphism $\hat{\bd{F}}$ becomes trivial,
$\pi_1^0\hat{\bd{F}}=\pi_1^0\id$\,, and the differential equations 
(\ref{diff eq open}) reduce to those in Ref.~\cite{Erler:2013xta} 
for the string product 
\begin{equation}
 \bd{A}|^0(s,t)\ =\ \sum_{n,m=0}^\infty s^m t^n \bd{A}^{(n)}_{n+m+1}|^0\,.
\end{equation}
It is also a generalization of that in Ref.~\cite{Erler:2016ybs},
in which a complete action of the open superstring field theory based on
the $A_\infty$ algebra was first constructed. The string products
with $A_\infty$ structure (\ref{A infty})
completely agree with those constructed in Ref.~\cite{Erler:2016ybs}
if we take the associative bosonic product,
$
 \bd{M}_B^{(0)}(s)=\bd{m}_2|^0+s\,\bd{m}_2|^2\,,
$
as an initial condition,\footnote{
The associative product $m_2$ can, for example, be defined by 
using the Witten's star product as
$
 m_2(\Phi_1,\Phi_2)=(-1)^{\textrm{deg}(\Phi_1)}\Phi_1*\Phi_2
$\,.
}
which is explicitly in Appendix~\ref{open product}.
Our construction method, however, allows us to take more general initial 
bosonic products, 
and provides more general 
superstring products which were not 
be able to be constructed before. 
This degree of freedom allows for the various open string field theories
realizing different triangulations of the moduli space,
which will be useful in analyzing specific problems.
The proof of cyclicity becomes easier in our construction,
which we give in Appendix \ref{app cyclicity}.

\subsection{S-matrix generating function and its evaluation}\label{open eval}

The tree-level S-matrix generating functional is obtained 
in a similar way to the heterotic string field theory:
\begin{equation}
 \langle S|\ =\ \langle\Omega|P_0\otimes\pi_1\bd{S}_{int}\,:\,
\mathcal{H}_0\otimes\mathcal{TH}_0\longrightarrow \mathbb{C}\,,
\end{equation}
with
\begin{equation}
 \bd{S}_{int}\ =\ \hat{\bd{P}}\bd{M}_{int}(\bd{I}-\bd{H}\bd{M}_{int})^{-1}\hat{\bd{P}}\,.
\end{equation}
Here, all the quantities are those acting on $\mathcal{TH}^{res}$\,,
which are similarly defined for the open superstring, 
\begin{equation}
 \bd{H}\ =\ \sum_{r,s=0}^\infty \id^{\otimes r}\otimes (-Q^+)\otimes P_0^{\otimes s}\,,\qquad
 \hat{\bd{P}}\ =\ \sum_{n=0}^\infty P_0^{\otimes n}\,,\qquad 
 \hat{\bd{I}}\ =\ \sum_{n=0}^\infty \id^{\otimes n}\,,
\end{equation}
using the open string homotopy operator
\begin{equation}
 Q^+\ =\ \frac{1}{L_0}b_0(1-P_0)\,,
\end{equation}
and satisfy the same relation as (\ref{basic prop}).
If we introduce two maps 
\begin{equation}
 \hat{\bd{i}}'\ =\ (\hat{\bd{I}}-\bd{H}\bd{M}_{int})^{-1}\hat{\bd{P}}\,,\qquad
\bd{\Sigma}\ =\ \bd{a}\hat{\bd{i}}'\,,
\end{equation}
the map $\bd{\Sigma}$ is related to $\bd{S}_{int}$ as
\begin{equation}
 \pi_1\bd{S}_{int}\ =\ \mathcal{G}P_0\pi_1\bd{\Sigma}\,,\quad
\langle S|\ =\ \langle\omega_l|\xi_0P_0\otimes P_0\pi_1\bd{\Sigma}\,,
\end{equation}
and satisfies the Dyson-Schwinger equation
\begin{equation}
\pi_1\bd{\Sigma}\ =\ \sum_{n=0}^\infty \pi_1\bd{A}_{n+2}
\Big((P_0\pi_1-\Delta\pi_1\bd{\Sigma})^{\otimes (n+2)}\Big)\,,
\end{equation}
where $\Delta=Q^+\mathcal{G}-\Xi\pi^1$\,.
These basic relations and the differential equations (\ref{diff eq open})
have the same form as those of the heterotic
string field theory, and therefore
it is easy to show that if we suppose 
that the two-parameter extension $\bd{\Sigma}(s,t)$ follows the extended
Dyson-Schwinger equation
\begin{equation}
\pi_1\bd{\Sigma}(s,t)\ =\ \sum_{n=0}^\infty \pi_1\bd{A}_{n+2}(s,t)
\Big((P_0\pi_1-\Delta(s,t)\pi_1\bd{\Sigma}(s,t))^{\otimes (n+2)}\Big)\,, 
\end{equation}
with 
\begin{equation}
\Delta(s,t)=Q^+\mathcal{G}(s,t)-t\Xi\pi^1\,,\qquad
\mathcal{G}(s,t)=\pi^0+(tX+s)\pi^1\,,
\end{equation}
$\bd{\Sigma}(s,t)$ commutes with $\bd{Q}$ and $\bd{\eta}$\,,
\begin{subequations}
 \begin{equation}
[\bd{Q}\,,\bd{\Sigma}(s,t)]\ =\ 0\,,\qquad
[\bd{\eta}\,,\bd{\Sigma}(s,t)]\ =\ 0\,,
\end{equation}
and satisfies
\begin{equation}
\partial_t\bd{\Sigma}(s,t)\ =\ [\bd{Q},\bd{\rho}(s,t)]\,,\qquad
\partial_s\bd{\Sigma}(s,t)\ =\ [\bd{\eta},\bd{\rho}(s,t)]\,,
\label{ders of Sigma A}
\end{equation}
 \end{subequations}
with certain $\bd{\rho}(s,t)$ determined in a manner similar to
that for the heterotic string given in Appendix~\ref{app 3}. 
%
The relations (\ref{ders of Sigma A}) lead to the same equation as
(\ref{der t der s}), from which we find that 
$\langle S(s,t)|=\langle\omega_l|\xi_0P_0\otimes P_0\pi_1\bd{\Sigma}(s,t)$
satisfies the equation
\begin{equation}
 \partial_t\langle S(s,t)|-\partial_s\langle S(s,t)|X_0\ =\ 0\,,
\end{equation}
with
\begin{equation}
 \langle S(s,t)|X_0\ =\ \sum_{n=0}^\infty
\langle S_{n+3}(s,t)|\Big(X_0\otimes\id_{n+2} + \id \otimes 
\sum_{r=0}^{n+1}\id_r\otimes X_0\otimes \id_{n-r+1}\Big)\,,
\end{equation}
except for the terms which vanish when acting on physical states.
Consequently, the physical S-matrix at the tree level can be rewritten as 
\begin{equation}
\langle S^{phys}|\ =\ 
\sum_{n=0}^\infty 
\sum_{p=0}^{n+1}\langle S^{(0)}|^{2(n-p+1)}(X_0)^p
\hat{\bd{\mathcal{P}}}\,,
\end{equation}
and agrees with that in the first-quantized formulation
since it has the form of integral of smooth functions (sections) 
with picture changed external states over the whole open-string moduli spaces.
%

\section{Summary and discussion}\label{summary}

In this paper, we have shown that the tree-level physical S-matrices derived
from the homotopy-algebra-based superstring field theories
reproduce those obtained in the first-quantized formulation. 
For the heterotic string field theory, 
the differential equations (\ref{diff eq}) for the (basic building blocks of) 
the string products $\bd{B}(s,t)$ play a key role.
Utilizing these differential equations we can eventually derive a sequence of 
decent equations (\ref{decent eq}) for the amplitudes. 
They allow us to rewrite the physical 
S-matrix in such a way that the equivalence to the one obtained in 
the first-quantized formulation is transparent.
The extension to the type II and open superstring field theories
has been straightforward since the key differential equations have 
essentially the same form.
The open superstring field theory considered in this paper is a generalization of
the previously constructed one and provides various theories realizing
different triangulations of the moduli space.

It is interesting to extend the proof to the S-matrix at the loop level.
In order to consider it, however, we need to extend 
the discussion to the one based on
the quantum (or loop) homotopy algebras \cite{Zwiebach:1992ie,Markl:1997bj,Munster:2011ij}.
For closed superstring field theories, 
it is related to the problem of how to avoid 
the unphysical spurious singularities \cite{Donagi:2013dua,Sen:2015hia}.
%
A consistent superstring field theory needs to be constructed to provide an algorithm 
that calculates the scattering amplitude as an integral over the moduli space 
that smoothly connects the contributions from various Feynman diagrams without 
hitting spurious singularities \cite{Erler:2017dgr}.
%
%
Some recent achievements 
\cite{Moosavian:2017fta,Moosavian:2017qsp,Moosavian:2017sev,Pius:2018pqr,Costello:2019fuh}
will help further progress.

There are many other interesting issues that can only be performed by means
of the string field theories \cite{deLacroix:2017lif}, and further studies are expected 
to be done.

\vspace{1cm}
\noindent
{\bf \large Acknowledgments}

The author would like to thank Tatsuya Sugimoto for collaboration
in the early stage of this work. He also would like to thank Jakub Vo\v{s}mera
for drawing his attention to Refs.~\cite{Erbin:2020eyc,Arvanitakis:2020rrk}.
He also would like to express his appreciation to organizers of ``Fundamental Aspects of 
String Theory'' ICTP-SAIFR, Sao Paulo 1-12 June 2020, in particular Nathan Berkovits, for providing
exciting atmosphere through the online workshop.
This work was supported in part by JSPS Grant-in-Aid for Scientific 
Research (C) Grant Number JP18K03645.

\vspace{1cm}
\appendix

\section{S-matrix via HPT}\label{app 1}

In this appendix, we derive the tree-level S-matrix of the heterotic 
string field theory (\ref{S gen func}) as the almost minimal model of
the $L_\infty$ algebra ($\mathcal{H}^{res},\Omega,\bd{L}$) 
by means of HPT.\footnote{
The minimal model usually considered \cite{Kajiura:2003ax,Kontsevich:1997vb} 
is the S-matrix naturally appeared in the light-cone gauge, 
in which all the unphysical model is gauged away \cite{Matsunaga:2019fnc}.}

Consider two chain complexes
$
(\mathcal{H}^{res},d=Q)$ and 
$
(\mathcal{H}_0,D=QP_0)$ with chain maps
\begin{equation}
p\ =\ P_0\,:\,\mathcal{H}^{res}\twoheadrightarrow\mathcal{H}_0\,,\qquad
i\ =\ P_0\,:\,\mathcal{H}_0\hookrightarrow\mathcal{H}^{res}\,,
\end{equation}
satisfying
\begin{equation}
 pi\ =\ \id_{\mathcal{H}_0}\ =\ P_0\,,\qquad
 ip\ =\ \id_{\mathcal{H}^{res}}+hd+dh\ =\
\id-Q^+Q-QQ^+\,.
\end{equation}
The chain complex $(\mathcal{H}_0,QP_0)$\,,
with the gauge conditions, 
defines
the relative BRST cohomology 
\cite{Figueroa-OFarrill:1988qnz,Figueroa-OFarrill:1988gxt,Lian:1989cy}.
In order to obtain the S-matrix (\ref{S matrix}) we first lift these
equivalence data to those acting on the corresponding symmetrized
tensor algebra defined by Eqs.~(\ref{homotopy ops}): 
$(\mathcal{SH}^{res},\bd{d}=\bd{Q})$ and 
$(\mathcal{SH}_0,\bd{D}=\bd{QP})$ with
\begin{align}
 \hat{\bd{p}}\ =\ \hat{\bd{P}}\,:\,\mathcal{SH}^{res}\twoheadrightarrow
\mathcal{SH}_0\,,\qquad
\hat{\bd{i}}\ =\ \hat{\bd{P}}\,:\, \mathcal{SH}_0\hookrightarrow
\mathcal{SH}^{res}\,,
\end{align}
satisfying
\begin{equation}
 \hat{\bd{p}}\hat{\bd{i}}\ =\ \hat{\bd{P}}\,,\qquad 
 \hat{\bd{i}}\hat{\bd{p}}\ =\ \hat{\bd{I}}+\bd{HQ}+\bd{QH}\,.
\end{equation}
Then, if we perturb $\bd{Q}$ by $\bd{L}_{int}$ so that
$
(\bd{Q}+\bd{L}_{int})^2=0$\,,
the homological perturbation lemma tells us that the equivalence data
are deformed as
\begin{align}
\bd{d}'\ =&\ \bd{Q}+\bd{L}_{int}\ =\ \bd{L}\,,\\
\hat{\bd{i}}'\ =&\ (\hat{\bd{I}}-\bd{HL}_{int})^{-1}\hat{\bd{P}}\,,\\
\hat{\bd{p}}'\ =&\ \hat{\bd{P}}(\hat{\bd{I}}-\bd{L}_{int}\bd{H})^{-1}\,,\\
\bd{D}'\ =&\ \bd{QP}+\hat{\bd{P}}\bd{L}_{int}(\hat{\bd{I}}-\bd{HL}_{int})^{-1}
\hat{\bd{P}}\ =\ \bd{QP}+\bd{S}_{int}\,,\\
\bd{h}'\ =&\ \bd{H}(\hat{\bd{I}}-\bd{L}_{int}\bd{H})^{-1}\ =\
(\hat{\bd{I}}-\bd{HL}_{int}^{-1})\bd{H}\,.
\end{align}
We confirm below that the almost minimal model $(\mathcal{SH}_0,\bd{D}')$
is actually $L_\infty$-quasi-isomorphic to $(\mathcal{SH}^{res},\bd{d}'=\bd{L})$ by 
showing that the deformed maps $\hat{\bd{i}}'$ and $\hat{\bd{p}}'$ are cohomomorphisms, 
and $\bd{D}'$ is a coderivation.

Note that coderivation $\bd{D}$ and cohomomorphism $\hat{\bd{F}}$ 
satisfy
\begin{align}
 \Delta\bd{D}\ =&\ (\bd{D}\otimes\id+\id\otimes\bd{D})\Delta\,,\\
 \Delta\hat{\bd{F}} =&\ (\hat{\bd{F}}\otimes\hat{\bd{F}})\Delta\,,
\end{align}
with respect to the coproduct $\Delta$\,.
%
On the other hand, the homotopy operator $\bd{H}$ defined by (\ref{def H}) 
itself does not have any good property under the coproduct $\Delta$\,. We have
\begin{align}
 \Delta\bd{H}\ 
=&\ \sum_{r,s,t,u=0}^\infty
\begin{pmatrix}
 r+t\\r
\end{pmatrix}
\begin{pmatrix}
 s+u\\u
\end{pmatrix}
\nonumber\\
&\hspace{0.5cm}\times
\Big\{
(-Q^+)\wedge\id^{\wedge r}\wedge P_0^{\wedge s}\otimes \id^{\wedge t}\wedge P_0^{\wedge u}
+\id^{\wedge r}\wedge P_0^{\wedge s}\otimes (-Q^+)\wedge\id^{\wedge t}\wedge P_0^{\wedge u}
\Big\}\Delta\,.
\end{align}
However, the complicated factor in front of the $\Delta$ in the right hand side 
becomes simple if it acts on 
$(\hat{\bd{P}}\otimes\bd{A}+\bd{A}\otimes\hat{\bd{P}})$
or $(\bd{H}\otimes\hat{\bd{I}}-\hat{\bd{I}}\otimes\bd{H})$
with any map $\bd{A}$ on $\mathcal{SH}^{res}$\,, and we find that
\begin{align}
&\ \sum_{r,s,t,u=0}^\infty
\begin{pmatrix}
 r+t\\r
\end{pmatrix}
\begin{pmatrix}
 s+u\\u
\end{pmatrix}
\nonumber\\
&\hspace{1cm}\times
\Big(
(-Q^+)\wedge\id^{\wedge r}\wedge P_0^{\wedge s}\otimes \id^{\wedge t}\wedge P_0^{\wedge u}
+\id^{\wedge r}\wedge P_0^{\wedge s}\otimes (-Q^+)\wedge\id^{\wedge t}\wedge P_0^{\wedge u}
\Big)\Big(\hat{\bd{P}}\otimes\bd{A}+\bd{A}\otimes\hat{\bd{P}}\Big)
\nonumber\\
&\ 
=\ \hat{\bd{P}}\otimes\bd{HA}+\bd{HA}\otimes\hat{\bd{P}}\,,
\label{formula on A P}
\end{align}
or
\begin{align}
&\ \sum_{r,s,t,u=0}^\infty
\begin{pmatrix}
 r+t\\r
\end{pmatrix}
\begin{pmatrix}
 s+u\\u
\end{pmatrix}
\nonumber\\
&\hspace{1cm}\times
\Big(
(-Q^+)\wedge\id^{\wedge r}\wedge P_0^{\wedge s}\otimes \id^{\wedge t}\wedge P_0^{\wedge u}
+\id^{\wedge r}\wedge P_0^{\wedge s}\otimes (-Q^+)\wedge\id^{\wedge t}\wedge P_0^{\wedge u}
\Big)\Big(\bd{H}\otimes\hat{\bd{I}}-\hat{\bd{I}}\otimes\bd{H}\Big)
\nonumber\\
&\  
=\ - \bd{H}\otimes\bd{H}\,.
\label{formula on H H}
\end{align}

The first relation (\ref{formula on A P}) is shown by using the formula
\begin{equation}
\sum_{u=0}^t
\begin{pmatrix}
 r+t-u\\r
\end{pmatrix}
\begin{pmatrix}
 s+u\\s
\end{pmatrix}\
=\ 
\begin{pmatrix}
 r+s+t+1\\ t
\end{pmatrix}\,,
\label{formula 1}
\end{equation}
which can be proven by mathematical induction with respect to $t$ as follows.
For $t=0$ (and $t=1$)\,,  Eq.~(\ref{formula 1}) is clearly true. 
Suppose that it holds for $t=t_0$\,, then we have
\begin{align}
 \sum_{u=0}^{t_0+1}
\begin{pmatrix}
 r+t_0+1-u\\r
\end{pmatrix}
\begin{pmatrix}
 s+u\\s
\end{pmatrix}\
=&\
\begin{pmatrix}
 r+t_0+1\\r
\end{pmatrix}
+ \sum_{u=1}^{t_0}
\begin{pmatrix}
 r+t_0+1-u\\r
\end{pmatrix}
\begin{pmatrix}
 s+u\\s
\end{pmatrix}\
+\begin{pmatrix}
  s+t_0+1\\s
 \end{pmatrix}
\nonumber\\
=&\
\sum_{u=0}^{t_0}\frac{r+t_0+1-u}{t_0+1}
\begin{pmatrix}
 r+t_0-u\\r
\end{pmatrix}
\begin{pmatrix}
 s+u\\s
\end{pmatrix}
\nonumber\\
&\
+
\sum_{u=1}^{t_0+1}\frac{s+u}{t_0+1}
\begin{pmatrix}
 r+t_0+1-u\\r
\end{pmatrix}
\begin{pmatrix}
 s+u-1\\s
\end{pmatrix}
\nonumber\\
=&\
\frac{r+s+t_0+2}{t_0+1}\sum_{u=0}^{t_0}
\begin{pmatrix}
 r+t_0-u\\r
\end{pmatrix}
\begin{pmatrix}
 s+u\\s
\end{pmatrix}
\nonumber\\
=&\
\begin{pmatrix}
 r+s+t_0+2\\t_0+1
\end{pmatrix}\,,
\label{proof binomial}
\end{align}
using the assumption of the induction in the last equality.
We used simple relations of the binomial coefficients
\begin{equation}
t
 \begin{pmatrix}
  t-1 \\ r
 \end{pmatrix}\
=\ (t-r)
\begin{pmatrix}
 t \\ r
\end{pmatrix}\
=\
(r+1)
\begin{pmatrix}
 t \\ r+1
\end{pmatrix}\,,\label{binomial 1}
\end{equation}
for $0\le r < t\in\mathbb{Z}$\,,
and the relation
\begin{align}
\begin{pmatrix}
 r+t_0+1-u\\r
\end{pmatrix}
\begin{pmatrix}
 s+u\\s
\end{pmatrix}\
=&\ 
\frac{r+t_0+1-u}{t_0+1}
\begin{pmatrix}
 r+t_0-u\\r
\end{pmatrix}
\begin{pmatrix}
 s+u\\s
\end{pmatrix}
\nonumber\\
&\
+ \frac{s+u}{t_0+1}
\begin{pmatrix}
 r+t_0+1-u\\r
\end{pmatrix}
\begin{pmatrix}
 s+u-1\\s
\end{pmatrix}\,,
\end{align}
for $1\le u\le t_0$\,, which can easily be confirmed using
(\ref{binomial 1}).
Since Eq.~(\ref{proof binomial}) shows that Eq.~(\ref{formula 1}) is true 
for $t=t_0+1$\,, Eq.~(\ref{formula 1}) is true by mathematical induction.

The second relation (\ref{formula on H H}) follows from the relation
\begin{equation}
\frac{1}{t!}\Big(\id^{\wedge t}\wedge P_0^{\wedge u}\Big)
\frac{1}{(t+u+1)!}\sum_{\alpha=0}^{t+u}
\Big(\id^{\wedge (t+u-\alpha)}\wedge P_0^{\wedge \alpha}\Big)\
=\
\frac{1}{(t+1)!}\sum_{\alpha=0}^t
\Big(\id^{\wedge (t-\alpha)}\wedge P_0^{\wedge (u+\alpha)}\Big)\,,
\label{relation 1}
\end{equation}
and the formula
\begin{align}
 \sum_{\alpha=0}^u
\begin{pmatrix}
 r+t-s-\alpha+1\\r-s
\end{pmatrix}
\begin{pmatrix}
 s+\alpha\\s
\end{pmatrix}
+
\sum_{\alpha=0}^s
\begin{pmatrix}
 r+t-u-\alpha+1\\t-u
\end{pmatrix}
\begin{pmatrix}
 u+\alpha\\u
\end{pmatrix}\
=\
\begin{pmatrix}
 r+t+2\\r+1
\end{pmatrix}\,,
\label{formula 2}
\end{align}
for $r\ge s$ and $t\ge u$\,.
In order to show the relation (\ref{relation 1}),
we first prove the formula
\begin{equation}
 \sum_{\beta=0}^u
\begin{pmatrix}
 u\\ \beta
\end{pmatrix}
\frac{(t+u-\alpha-\beta)!}{(t-\alpha)!}\frac{(\alpha+\beta)!}{\alpha!}\
=\
\frac{(t+u+1)!}{(t+1)!}\,,
\label{sub formula}
\end{equation}
by mathematical induction with respect to $u$ as follows.
For $u=0$\,, it clearly holds. Suppose that it holds for $u=u_0$\,, then
\begin{align}
&\
 \sum_{\beta=0}^{u_0+1}
\begin{pmatrix}
 u_0+1\\ \beta
\end{pmatrix}
\frac{(t+u_o+1-\alpha-\beta)!}{(t-\alpha)!}\frac{(\alpha+\beta)!}{\alpha!}
\nonumber\\ 
&\
=\
\frac{(t+u_0+1-\alpha)!}{(t-\alpha)!}
+ \sum_{\beta=1}^{u_0}
\begin{pmatrix}
 u_0+1\\ \beta
\end{pmatrix}
\frac{(t+u_0+1-\alpha-\beta)!}{(t-\alpha)!}\frac{(\alpha+\beta)!}{\alpha!}
+\frac{(\alpha+u_0+1)!}{\alpha!}
\nonumber\\
&\
=\
\sum_{\beta=0}^{u_0}
\begin{pmatrix}
 u_0\\ \beta
\end{pmatrix}
\frac{(t+u_0+1-\alpha-\beta)}{(t-\alpha)!}
\frac{(\alpha+\beta)!}{\alpha!}
+
\sum_{\beta=1}^{u_0+1}
\begin{pmatrix}
 u_0\\ \beta-1
\end{pmatrix}
\frac{(t+u_0+1-\alpha-\beta)!}{(t-\alpha)!}
\frac{(\alpha+\beta)!}{\alpha!}
\nonumber\\
&\
=\
(t+u_0+2)\sum_{\beta=0}^{u_0}
\begin{pmatrix}
 u_0\\ \beta
\end{pmatrix}
\frac{(t+u_0-\alpha-\beta)!}{(t-\alpha)!}\frac{(\alpha+\beta)!}{\alpha!}
\nonumber\\
&\
=\
\frac{(t+u_0+2)!}{(t+1)!}\,,
\end{align}
by using the relation
\begin{equation}
\begin{pmatrix}
 u_0+1\\ \beta
\end{pmatrix}\
=\
\begin{pmatrix}
 u_0\\ \beta
\end{pmatrix}
+
\begin{pmatrix}
 u_0\\ \beta - 1
\end{pmatrix}\,,
\label{comb formula}
\end{equation}
for $1\le \beta \le u_0$\,.
This shows that (\ref{sub formula}) holds for $u=u_0+1$\,,
and hence the formula (\ref{sub formula})  
is proven by mathematical induction.
Using this formula (\ref{sub formula}), 
the left hand side of the relation (\ref{relation 1}) can be
calculated by noting that $\frac{1}{t!}\id^{\wedge t}$ is 
the identity on $(\mathcal{H}^{res})^{\wedge t}$\,:
\begin{align}
&\
\frac{1}{t!}\Big(\id^{\wedge t}\wedge P_0^{\wedge u}\Big)
\frac{1}{(t+u+1)!}\sum_{\alpha=0}^{t+u}
\Big(\id^{\wedge (t+u-\alpha)}\wedge P_0^{\wedge \alpha}\Big)\
\nonumber\\
&\
=\
 \frac{1}{(t+u+1)!}
\sum_{\beta=0}^u\sum_{\alpha=\beta}^{t+\beta}
\begin{pmatrix}
 u\\ \beta
\end{pmatrix}
\frac{(t+u-\alpha)!}{(t+\beta-\alpha)!}
\frac{\alpha!}{(\alpha-\beta)!}\,
\id^{\wedge (t+\beta-\alpha)}\wedge P_0^{\wedge (u+\alpha-\beta)}
\nonumber\\
&\
=\
 \frac{1}{(t+u+1)!}
\sum_{\alpha=0}^t
\Bigg(
\sum_{\beta=0}^u
\begin{pmatrix}
 u\\ \beta
\end{pmatrix}
\frac{(t+u-\alpha-\beta)!}{(t-\alpha)}\frac{(\alpha+\beta)!}{\alpha!}
\Bigg)
\id^{\wedge (t-\alpha)}\wedge P_0^{\wedge (u+\alpha)}
\nonumber\\
&\
=\
\frac{1}{(t+1)!}
\sum_{\alpha=0}^t
\id^{\wedge (t-\alpha)}\wedge P_0^{\wedge (u+\alpha)}\,.
\end{align}
The proof of the formula (\ref{formula 2}) is the following.
We assume that $s\ge u$\,, which is possible without loss of generality
since the formula
is symmetric with respect to $s$ and $u$\,,
and use the mathematical induction with respect to $u$\,.
First, rewrite the formula as
\begin{equation}
 \sum_{\alpha=0}^u
\begin{pmatrix}
 r+t-s-\alpha+1\\r-s
\end{pmatrix}
\begin{pmatrix}
 s+\alpha\\s
\end{pmatrix}
+
\sum_{\alpha=0}^s
\begin{pmatrix}
 r+t-s-u+\alpha+1\\r-s+\alpha+1
\end{pmatrix}
\begin{pmatrix}
 s+u-\alpha\\s-\alpha
\end{pmatrix}\
=\
\begin{pmatrix}
 r+t+2\\r+1
\end{pmatrix}\,.
\label{formula 2 rev}
\end{equation}
For $u=0$\,, the left hand side is calculated as
\begin{align}
&\
 \begin{pmatrix}
  r+t-s+1\\ r-s
 \end{pmatrix}
+ \sum_{\alpha=0}^s
\begin{pmatrix}
r+t-s+\alpha+1\\r-s+\alpha+1 
\end{pmatrix}
\nonumber\\
&\
=\
 \begin{pmatrix}
  r+t-s+1\\ r-s
 \end{pmatrix}
+
\begin{pmatrix}
 r+t-s+1\\r-s+1
\end{pmatrix}
+ \sum_{\alpha=1}^s
\begin{pmatrix}
r+t-s+\alpha+1\\r-s+\alpha+1 
\end{pmatrix}
\nonumber\\
&\
=\ \cdots
\nonumber\\
&\
=\ 
\begin{pmatrix}
 r+t-s+k+1\\r-s+k
\end{pmatrix}
+
\begin{pmatrix}
 r+t-s+k+1\\r-s+k+1
\end{pmatrix}
+ \sum_{\alpha=k+1}^s
\begin{pmatrix}
r+t-s+\alpha+1\\r-s+\alpha+1 
\end{pmatrix}
\nonumber\\
&\
=\ \cdots
\nonumber\\
&\
=\
\begin{pmatrix}
r+t+1\\r
\end{pmatrix}
+
\begin{pmatrix}
 r+t+1\\r+1
\end{pmatrix}\
=\
\begin{pmatrix}
 r+t+2\\r+1
\end{pmatrix}\,,
\end{align}
by using the relation (\ref{comb formula}) repeatedly.
Then, suppose that it is true for $u=u_0$\,. 
The left hand side of (\ref{formula 2 rev})
for $u=u_0+1$ can be calculated as
\begin{align}
 \sum_{\alpha=0}^{u_0+1}
&\
\begin{pmatrix}
 r+t-s-\alpha+1\\r-s
\end{pmatrix}
\begin{pmatrix}
 s+\alpha\\s
\end{pmatrix}
+
\sum_{\alpha=0}^s
\begin{pmatrix}
 r+t-s-u_0+\alpha\\r-s+\alpha+1
\end{pmatrix}
\begin{pmatrix}
 s+u_0-\alpha+1\\s-\alpha
\end{pmatrix}
\nonumber\\
=&\
 \sum_{\alpha=0}^{u_0}
\begin{pmatrix}
 r+t-s-\alpha+1\\r-s
\end{pmatrix}
\begin{pmatrix}
 s+\alpha\\s
\end{pmatrix}
+
\begin{pmatrix}
 r+t-s-u_0\\r-s
\end{pmatrix}
\begin{pmatrix}
 s+u_0+1\\s
\end{pmatrix}
\nonumber\\
&\
+
\begin{pmatrix}
 r+t-s-u_0\\r-s+1
\end{pmatrix}
\begin{pmatrix}
 s+u_0+1\\s
\end{pmatrix}
+
\sum_{\alpha=1}^s
\begin{pmatrix}
 r+t-s-u_0+\alpha\\r-s+\alpha+1
\end{pmatrix}
\begin{pmatrix}
 s+u_0-\alpha+1\\s-\alpha
\end{pmatrix}
\nonumber\\
=&\
 \sum_{\alpha=0}^{u_0}
\begin{pmatrix}
 r+t-s-\alpha+1\\r-s
\end{pmatrix}
\begin{pmatrix}
 s+\alpha\\s
\end{pmatrix}\
\nonumber\\
&\
+
\begin{pmatrix}
 r+t-s-u_0+1\\r-s+1
\end{pmatrix}
\begin{pmatrix}
 s+u_0+1\\s
\end{pmatrix}
+
\sum_{\alpha=1}^s
\begin{pmatrix}
 r+t-s-u_0+\alpha\\r-s+\alpha+1
\end{pmatrix}
\begin{pmatrix}
 s+u_0-\alpha+1\\s-\alpha
\end{pmatrix}
\nonumber\\
=&\
 \sum_{\alpha=0}^{u_0}
\begin{pmatrix}
 r+t-s-\alpha+1\\r-s
\end{pmatrix}
\begin{pmatrix}
 s+\alpha\\s
\end{pmatrix}
\nonumber\\
&\
+
\begin{pmatrix}
 r+t-s-u_0+1\\r-s+1
\end{pmatrix}
\begin{pmatrix}
 s+u_0\\s
\end{pmatrix}
+
\begin{pmatrix}
 r+t-s-u_0+1\\r-s+1
\end{pmatrix}
\begin{pmatrix}
 s+u_0\\s-1
\end{pmatrix}
\nonumber\\
&\
+
\begin{pmatrix}
 r+t-s-u_0+1\\r-s+2
\end{pmatrix}
\begin{pmatrix}
 s+u_0\\s-1
\end{pmatrix}
+
\sum_{\alpha=2}^s
\begin{pmatrix}
 r+t-s-u_0+\alpha\\r-s+\alpha+1
\end{pmatrix}
\begin{pmatrix}
 s+u_0-\alpha+1\\s-\alpha
\end{pmatrix}
\nonumber
\end{align}
\begin{align}
=&\
 \sum_{\alpha=0}^{u_0}
\begin{pmatrix}
 r+t-s-\alpha+1\\r-s
\end{pmatrix}
\begin{pmatrix}
 s+\alpha\\s
\end{pmatrix}
+
\begin{pmatrix}
 r+t-s-u_0+1\\r-s+1
\end{pmatrix}
\begin{pmatrix}
 s+u_0\\s
\end{pmatrix}
\nonumber\\
&\
+
\begin{pmatrix}
 r+t-s-u_0+2\\r-s+2
\end{pmatrix}
\begin{pmatrix}
 s+u_0\\s-1
\end{pmatrix}
\nonumber\\
&\
+
\sum_{\alpha=2}^s
\begin{pmatrix}
 r+t-s-u_0+\alpha\\r-s+\alpha+1
\end{pmatrix}
\begin{pmatrix}
 s+u_0-\alpha+1\\s-\alpha
\end{pmatrix}
\nonumber\\
=&\ 
 \sum_{\alpha=0}^{u_0}
\begin{pmatrix}
 r+t-s-\alpha+1\\r-s
\end{pmatrix}
\begin{pmatrix}
 s+\alpha\\s
\end{pmatrix}
+
\begin{pmatrix}
 r+t-s-u_0+1\\r-s+1
\end{pmatrix}
\begin{pmatrix}
 s+u_0\\s
\end{pmatrix}
\nonumber\\
&\
+
\begin{pmatrix}
 r+t-s-u_0+2\\r-s+2
\end{pmatrix}
\begin{pmatrix}
 s+u_0-1\\s-1
\end{pmatrix}
+
\begin{pmatrix}
 r+t-s-u_0+2\\r-s+2
\end{pmatrix}
\begin{pmatrix}
 s+u_0-1\\s-2
\end{pmatrix}
\nonumber\\
&\
+
\begin{pmatrix}
 r+t-s-u_0+2\\r-s+3
\end{pmatrix}
\begin{pmatrix}
 s+u_0-1\\s-2
\end{pmatrix}
+
\sum_{\alpha=3}^s
\begin{pmatrix}
 r+t-s-u_0+\alpha\\r-s+\alpha+1
\end{pmatrix}
\begin{pmatrix}
 s+u_0-\alpha+1\\s-\alpha
\end{pmatrix}
\nonumber\\
=&\ \cdots
\nonumber\\
=&\
 \sum_{\alpha=0}^{u_0}
\begin{pmatrix}
 r+t-s-\alpha+1\\r-s
\end{pmatrix}
\begin{pmatrix}
 s+\alpha\\s
\end{pmatrix}
+
\sum_{\alpha=0}^{k-1}
\begin{pmatrix}
 r+t-s-u_0+\alpha+1\\r-s+\alpha+1
\end{pmatrix}
\begin{pmatrix}
 s+u_0-\alpha\\s-\alpha
\end{pmatrix}
\nonumber\\
&\
+
\begin{pmatrix}
 r+t-s-u_0+k\\r-s+k
\end{pmatrix}
\begin{pmatrix}
 s+u_0-k+2\\s-k+1
\end{pmatrix}
\nonumber\\
&\
+
\sum_{\alpha=k}^s
\begin{pmatrix}
 r+t-s-u_0+\alpha\\r-s+\alpha+1
\end{pmatrix}
\begin{pmatrix}
 s+u_0-\alpha+1\\s-\alpha
\end{pmatrix}
\nonumber\\
=&\ \cdots
\nonumber\\
=&\
 \sum_{\alpha=0}^{u_0}
\begin{pmatrix}
 r+t-s-\alpha+1\\r-s
\end{pmatrix}
\begin{pmatrix}
 s+\alpha\\s
\end{pmatrix}
+
\sum_{\alpha=0}^{s-1}
\begin{pmatrix}
 r+t-s-u_0+\alpha+1\\r-s+\alpha+1
\end{pmatrix}
\begin{pmatrix}
 s+u_0-\alpha\\s-\alpha
\end{pmatrix}
\nonumber\\
&\
+
\begin{pmatrix}
 r+t-u_0\\r
\end{pmatrix}
+
\begin{pmatrix}
 r+t-u_0\\r+1
\end{pmatrix}
\nonumber\\
=&\
 \sum_{\alpha=0}^{u_0}
\begin{pmatrix}
 r+t-s-\alpha+1\\r-s
\end{pmatrix}
\begin{pmatrix}
 s+\alpha\\s
\end{pmatrix}
+
\sum_{\alpha=0}^{s}
\begin{pmatrix}
 r+t-s-u_0+\alpha+1\\r-s+\alpha+1
\end{pmatrix}
\begin{pmatrix}
 s+u_0-\alpha\\s-\alpha
\end{pmatrix}
\nonumber\\
=&\
\begin{pmatrix}
 r+t+2\\r+1
\end{pmatrix}\,.
\end{align}
The last equality follows from the assumption of the induction.
Hence, the formula (\ref{formula 2}) is true for ${}^\forall u$
by mathematical induction.

Using these properties of $\bd{H}$\,,
we can show, order by order in the power of $\bd{H}$\,, 
that the linear map on $\mathcal{SH}_0$\,,
\begin{align}
\hat{\bd{i}}'\ =&\ (\hat{\bd{I}}-\bd{H}\bd{L}_{int})^{-1}\hat{\bd{P}}\
=\ \sum_{n=0}^\infty (\bd{H}\bd{L}_{int})^n\hat{\bd{P}}\,,
\label{Def}
\end{align}
satisfies the property of the cohomomorphism
\begin{equation}
 \Delta\hat{\bd{i}}'\ =\ (\hat{\bd{i}}'\otimes\hat{\bd{i}}')\Delta\,.
\label{cohomo}
\end{equation}
At $\mathcal{O}((\bf{H})^0)$\,, Eq.~(\ref{cohomo}) reduces to 
$\Delta\hat{\bd{P}}=(\hat{\bd{P}}\otimes\hat{\bd{P}})\Delta$\,, which
follows from the definition (\ref{def P}) of $\hat{\bd{P}}$\,.
For $\mathcal{O}((\bd{H})^n)$ with $n\ge1$\,, we use the mathematical induction.
If we note that
\begin{equation}
 \Delta\bd{L}_{int}\hat{\bd{P}}\ 
=\ (\hat{\bd{P}}\otimes\bd{L}_{int}\hat{\bd{P}}
+ \bd{L}_{int}\hat{\bd{P}}\otimes\hat{\bd{P}})\Delta\,,
\end{equation}
and the factor in the right hand side
has the form mentioned above, we can see that
$\mathcal{O}(\bd{H})$  of Eq.~(\ref{cohomo}),
\begin{equation}
 \Delta(\bd{H}\bd{L}_{int}\hat{\bd{P}})\ 
=\ (\hat{\bd{P}}\otimes\bd{H}\bd{L}_{int}\hat{\bd{P}}
+\bd{H}\bd{L}_{int}\hat{\bd{P}}\otimes\hat{\bd{P}})\Delta\,,
\end{equation}
also holds thanks to the relation (\ref{formula on A P}).
If we suppose that $\mathcal{O}((\bd{H})^{n_0})$ of (\ref{cohomo}),
\begin{equation}
 \Delta\big((\bd{H}\bd{L}_{int})^{n_0}\hat{\bd{P}}\big)\ =\
\Big(\sum_{m=0}^{n_0}\big((\bd{H}\bd{L}_{int})^{n_0-m}\hat{\bd{P}}
\otimes (\bd{H}\bd{L}_{int})^{m}\hat{\bd{P}}\big)\Big)\Delta\,,
\end{equation}
holds, then from
\begin{align}
\Delta\bd{L}_{int}&(\bd{H}\bd{L}_{int})^{n_0}\hat{\bd{P}}\ 
\nonumber\\
=&\
\Big(\hat{\bd{P}}\otimes\bd{L}_{int}(\bd{H}\bd{L}_{int})^{n_0}\hat{\bd{P}}
+ \bd{L}_{int}(\bd{H}\bd{L}_{int})^{n_0}\hat{\bd{P}}\otimes \hat{\bd{P}}
\nonumber\\
&\
+ (\bd{H}\otimes\hat{\bd{I}}-\hat{\bd{I}}\otimes\bd{H}) 
\sum_{m=0}^{n_0-1}\big(
\bd{L}_{int}(\bd{H}\bd{L}_{int})^{n_0-m-1}\hat{\bd{P}}
\otimes\bd{L}_{int}(\bd{H}\bd{L}_{int})^{m}\hat{\bd{P}}
\big)
\Big)\Delta\,,
\end{align}
the relation at $\mathcal{O}((\bd{H})^{n_0+1})$\,,
\begin{equation}
 \Delta\big((\bd{H}\bd{L}_{int})^{n_0+1}\hat{\bd{P}}\big)\ =\
\Big(\sum_{m=0}^{n_0+1}\big((\bd{H}\bd{L}_{int})^{n_0-m+1}\hat{\bd{P}}
\otimes (\bd{H}\bd{L}_{int})^{m}\hat{\bd{P}}\big)\Big)\Delta\,,
\end{equation}
holds again thanks to the relations (\ref{formula on A P}) 
and (\ref{formula on H H}).
Hence, Eq.~(\ref{cohomo}) holds at any order,
$\mathcal{O}((\bd{H})^n)$ with ${}^\forall n$\,,
due to the mathematical induction, and thus,
the map $\hat{\bd{i}}'$ is a cohomomorphism.
Note that the cohomomorphism $\hat{\bd{i}}'$
is cyclic with respect to $\Omega$ in the sense that it satisfies
\begin{equation}
\langle\Omega|\pi_1\hat{\bd{i}}'\otimes\pi_1\hat{\bd{i}}'\
=\ \langle\Omega|P_0\otimes P_0\,.
\label{cyclicity of i}
\end{equation}
This follows from Eqs.~(\ref{prop Q+}), and
the fact that the symplectic form $\langle\Omega|$ compatible
with the decomposition (\ref{HK}) satisfies
\begin{equation}
\langle\Omega|\id\otimes Q^+\ =\ \langle\Omega|Q^+\otimes\id\,,\qquad
\langle\Omega|\id\otimes P_0\ =\ \langle\Omega|P_0\otimes\id\,. 
\end{equation}

We can also show that $\hat{\bd{p}}'$ is a cohomomorphism
using the similar relations to
(\ref{formula on A P}) and (\ref{formula on H H}), 
\begin{align}
\Big(\hat{\bd{P}}\otimes\bd{A}+\bd{A}\otimes\hat{\bd{P}}\Big)
&\
\sum_{r,s,t,u=0}^\infty
\begin{pmatrix}
 r+t\\r
\end{pmatrix}
\begin{pmatrix}
 s+u\\u
\end{pmatrix}
\nonumber\\
&\hspace{1cm}\times
\Big(
(-Q^+)\wedge\id^{\wedge r}\wedge P_0^{\wedge s}\otimes \id^{\wedge t}\wedge P_0^{\wedge u}
+\id^{\wedge r}\wedge P_0^{\wedge s}\otimes (-Q^+)\wedge\id^{\wedge t}\wedge P_0^{\wedge u}
\Big)
\nonumber\\
&\ 
=\ \hat{\bd{P}}\otimes\bd{AH}+\bd{AH}\otimes\hat{\bd{P}}\,,
\label{formula on A P 2}
\end{align}
and
\begin{align}
\Big(\bd{H}\otimes\hat{\bd{I}}-\hat{\bd{I}}\otimes\bd{H}\Big)
&\ \sum_{r,s,t,u=0}^\infty
\begin{pmatrix}
 r+t\\r
\end{pmatrix}
\begin{pmatrix}
 s+u\\u
\end{pmatrix}
\nonumber\\
&\hspace{1cm}\times
\Big(
(-Q^+)\wedge\id^{\wedge r}\wedge P_0^{\wedge s}\otimes \id^{\wedge t}\wedge P_0^{\wedge u}
+\id^{\wedge r}\wedge P_0^{\wedge s}\otimes (-Q^+)\wedge\id^{\wedge t}\wedge P_0^{\wedge u}
\Big)
\nonumber\\
&\  
=\  \bd{H}\otimes\bd{H}\,,
\label{formula on H H 2}
\end{align}
respectively, obtained by reversing the order of the factors.
We omit the explicit proof since it is parallel to that for $\hat{\bd{i}}'$\,.

\section{Derivation of (\ref{s gen func})}\label{app S matrix}

Note that the action (\ref{master action}) can also be written as a WZW-like form
\begin{equation}
 I[\Phi]\ =\ \int_0^1 dt\, \Omega(\partial_t\Phi(t),\pi_1\bd{L}(e^{\wedge\Phi(t)}))\,,
\label{WZW-like action}
\end{equation}
where $\Phi(t)$ is a one-parameter, $t\in[0,1]$\,,
extension of the string field $\Phi$ satisfying
$\Phi(1)=\Phi$ and $\Phi(0)=0$\,. 
This can easily be seen as follows.
First, note that
the action (\ref{WZW-like action}) agrees with the action (\ref{master action})
if we take $\Phi(t)=t\Phi$\,.
On the other hand, it is easy to show that an arbitrary variation of the integrand
becomes total $t$-derivative,
\begin{equation}
 \delta\Omega(\partial_t\Phi(t),\pi_1\bd{L}(e^{\wedge\Phi(t)}))\
=\ 
 \partial_t\Omega(\delta\Phi(t),\pi_1\bd{L}(e^{\wedge\Phi(t)}))\,.
\end{equation}
Thus, the action (\ref{master action}) is not changed by deforming
the $t$-dependence of $\Phi(t)$ as long as keeping the boundary condition:
$\delta\Phi(0)=\delta\Phi(1)=0$\,.
Then, we extend the classical solution (\ref{sol 1}) as
\begin{equation}
 \phi_{cl}(t)\ =\ \pi_1(\hat{\bd{I}}-\bd{H}\bd{L}_{int})^{-1}
\hat{\bd{P}}(e^{\wedge t\phi_0})\,,
\end{equation}
and evaluate the action as
\begin{align}
 I[\phi_{cl}]\ =&\ 
\int_0^1dt\,\Omega(\pi_1(\hat{\bd{I}}-\bd{H}\bd{L}_{int})^{-1}\hat{\bd{P}}
(\phi_0\wedge e^{\wedge t\phi_0}),\pi_1\bd{L}(\hat{\bd{I}}-\bd{H}\bd{L}_{int})^{-1}\hat{\bd{P}}
(e^{\wedge t\phi_0}))
\nonumber\\
=&\
\int_0^1dt\,\langle\Omega|\,\phi_0\otimes\pi_1\bd{S}(e^{\wedge t\phi_0})\,,
\label{S in app}
\end{align}
using the bilinear map representation and the relation 
\begin{equation}
(\hat{\bd{I}}-\bd{H}\bd{L}_{int})\bd{L}(\hat{\bd{I}}-\bd{H}\bd{L}_{int})^{-1}\
=\ \bd{Q}+\hat{\bd{P}}\bd{L}_{int}(\hat{\bd{I}}-\bd{H}\bd{L}_{int})\,,
\end{equation}
that follows from (\ref{prop 1}), (\ref{cyclicity of i}), and
$\bd{L}^2=(\bd{Q}+\bd{L}_{int})^2=0$\,.
The S-matrix generating functional (\ref{s gen func}) is obtained by 
explicitly carrying out the $t$-integration. 

\section{Proof of (\ref{commutativity})}\label{app 2}

In order to explicitly investigate $\bd{\Sigma}(s,t)$\,, it is necessary to
expand it in the number of inputs as (\ref{Sigma expanded}).
The generalized Dyson-Schwinger equation (\ref{general recursive Sigma})
is then written as
\begin{align}
  \pi_1\bd{\Sigma}_{n+2}(s,t)\ 
=&\
\sum_{m=0}^\infty
\pi_1\bd{B}_{m+2}(s,t)\bigg(\frac{1}{(m+2)!}
\Big(P_0\pi_1
-\Delta(s,t)\sum_{l=0}^{n-m-1}\pi_1\bd{\Sigma}_{l+2}(s,t)\Big)^{\wedge(m+2)}
\bigg)\pi_{n+2}\,,
\nonumber\\
\equiv&\ 
\sum_{m=0}^{n}\pi_1\bd{B}_{m+2}(s,t)
\Big(D_{m+2}(s,t)\Big)\pi_{n+2}\,,
\label{ext sigma n+2}
\end{align}
where we defined
\begin{equation}
D_M(s,t)\ \equiv\ \frac{1}{M!}\Big(P_0\pi_1
-\Delta(s,t)\pi_1\bd{\Sigma}_{}(s,t)\Big)^{\wedge M}\,,
\end{equation}
for notational simplicity.
Note that Eq.~(\ref{ext sigma n+2}) is a recurrence relation 
whose right hand side includes only $\bd{\Sigma}_{l+2}(s,t)$ with $l<n$\,.
It determines all the $\bd{\Sigma}_{n+2}(s,t)$ $(n\ge0)$
recursively from the relation for $n=0$\,:
$\pi_1\bd{\Sigma}_2(s,t)=\pi_1\bd{B}_2(s,t)P_2$\,.
Then,
Eq.~(\ref{commutativity}) can be proved by complete induction 
with respect to the number of external strings. 
For $n=0$\,, we have 
\begin{alignat}{3}
 [\bd{Q},\pi_1\bd{\Sigma}_2(s,t)]\ =&\
[\bd{Q},\pi_1\bd{B}_2(s,t)P_2]\ =&\ 0\,,\\
 [\bd{\eta},\pi_1\bd{\Sigma}_2(s,t)]\ =&\ 
[\bd{\eta},\pi_1\bd{B}_2(s,t)P_2]\ =&\ 0\,,
\end{alignat}
which follow
from 
the relations for $\bd{B}_2(s,t)$
in Eqs.~(\ref{L infty fixed ex}).
Next, suppose that
\begin{equation}
 [\bd{Q},\pi_1\bd{\Sigma}_{l+2}(s,t)]\
=\ [\bd{\eta},\pi_1\bd{\Sigma}_{l+2}(s,t)]\ =\ 0\,,
\end{equation} 
for $0\le l\le n-1$\,.
Then, we find from Eq.~(\ref{ext sigma n+2}) that
\begin{align}
&\ [\bd{Q},\pi_1\bd{\Sigma}_{n+2}(s,t)]
\nonumber\\
&\
=\ - \sum_{l=0}^{n-1}\pi_1\bd{B}_{l+2}(s,t)
\bigg(D_{l+1}(s,t)\wedge
\pi(s)\pi_1\sum_{m=0}^{n-l-1}\bd{B}_{m+2}(s,t)\Big(
D_{m+2}(s,t)\Big)
\bigg)\pi_{n+2}
\nonumber\\
&\hspace{10mm}
+ \sum_{m=0}^{n-1}\pi_1\bd{B}_{m+2}(s,t)
\bigg(D_{m+1}(s,t)\wedge
\pi(s)
\pi_1\bd{\Sigma}(s,t)
\bigg)\pi_{n+2}
\nonumber\\
&\
=\ 0\,,
\end{align}
using the relation (\ref{Q B alt form}) and
\begin{equation}
 [Q\,,\Delta(s,t)]\ =\ \pi(s)-P_0\mathcal{G}(s,t)\,.
\end{equation}
It was also used the fact that
the internal states are generically off-shell.
Hence, from the principle of mathematical induction,
we can conclude that $[\bd{Q},\pi_1\bd{\Sigma}(s,t)]=0$\,.

Similarly, we find from Eq.~(\ref{ext sigma n+2}) that
\begin{align}
&\ [\bd{\eta},\pi_1\bd{\Sigma}_{n+2}(s,t)]
\nonumber\\
&\
=\ - \sum_{l=0}^{n-1}\pi_1\bd{B}_{l+2}(s,t)
\bigg(D_{l+1}(s,t)\wedge
t\pi^1_1\sum_{m=0}^{n-l-1}\bd{B}_{m+2}(s,t)\Big(
D_{m+2}(s,t)\Big)\bigg)\pi_{n+2}
\nonumber\\
&\hspace{10mm}
+ \sum_{m=0}^{n-1}\pi_1\bd{B}_{m+2}(s,t)
\bigg(D_{m+1}(s,t)\wedge
t\pi^1_1\bd{\Sigma}(s,t)\bigg)\pi_{n+2}
\nonumber\\
&\
=\ 0\,,
\end{align}
using the relation (\ref{Q eta alt form}) and 
\begin{equation}
 [\eta\,,\Delta(s,t)]\ =\ -t\pi^1\,,
\end{equation}
with the off-shell-ness
of the internal states.
Hence, it can also be
concluded that $[\bd{\eta},\pi_1\bd{\Sigma}(s,t)]=0$
from the principle of mathematical induction.

\section{Proof of (\ref{ders of Sigma})}\label{app 3}

The proof of (\ref{ders of Sigma})
is given by complete induction with respect to
the number of inputs. For $\bd{\Sigma}_2(s,t)=\bd{B}_2(s,t)P_2$\,,
we have 
\begin{equation}
 \pi_1\partial_t\bd{\Sigma}_2(s,t)\ =\ [\bd{Q},\pi_1\bd{\rho}_2(s,t)]\,,
\end{equation}
with $\pi_1\bd{\rho}_2(s,t)=\pi_1\bd{\lambda}_2(s,t)P_2$
using (\ref{diff t alt}) with $n=0$\,:
\begin{equation}
 \partial_t\bd{B}_2(s,t)\ =\ [\bd{Q}\,,\bd{\lambda}_2(s,t)]\,.
\end{equation}
Next, for 
\begin{align}
 \bd{\Sigma}_3(s,t)\ =&\ \bd{B}_3(s,t)P_3
-\bd{B}_2(s,t)\left(\Delta(s,t)\pi_1
\bd{\Sigma}_2(s,t)
\wedge P_0\pi_1\right)\,,
\end{align}
we can find that
\begin{equation} 
\pi_1\partial_t\bd{\Sigma}_3(s,t)\ =\ [\bd{Q},\pi_1\bd{\rho}_3(s,t)]\,,
\end{equation}
with
\begin{align}
\pi_1\bd{\rho}_3(s,t)\ =&\ 
\pi_1\bd{\lambda}_3(s,t)P_3
- \pi_1\bd{\lambda}_2(s,t)\Big(\Delta(s,t)\pi_1\bd{\Sigma}_2(s,t)\wedge P_0\pi_1\Big)
\nonumber\\
&\hspace{10mm}
- \pi_1\bd{B}_2(s,t)\Big(\big[\Delta(s,t)\pi_1\bd{\rho}_2(s,t)
+Q^+\Xi\pi_1^1\bd{\Sigma}_2(s,t)\big]\wedge P_0\pi_1\Big)\,,
\end{align}
using (\ref{diff t alt}) with $n=1$\,, 
\begin{align}
 \partial_t\bd{B}_3(s,t)\ =&\ [\bd{Q}\,,\bd{\lambda}_3(s,t)]
\nonumber\\
&\
+ \bd{B}_2(s,t)\Big(\pi(s)\pi_1\bd{\lambda}_2(s,t)\wedge\id_1\Big)
- \bd{\lambda}_2(s,t)\Big(\pi(s)\pi_1\bd{B}_2(s,t)\wedge\id_1\Big)\,,
\end{align}
and
\begin{equation}
\partial_t\Delta(s,t)\ =\ - [Q\,, Q^+\Xi\pi^1] - \Xi P_0\pi^1\,.
\end{equation}
It was also used the off-shell-ness of the internal state.
Next, suppose that
\begin{align}
\pi_1\partial_t\bd{\Sigma}_{l+2}(s,t)\ =&\ [\bd{Q},\pi_1\bd{\rho}_{l+2}(s,t)]\,,\\
\pi_1\bd{\rho}_{l+2}(s,t)\ =&\ 
\sum_{m=0}^l\pi_1\bd{\lambda}_{m+2}(s,t)\big(
D_{m+2}(s,t)\big)\pi_{l+2}
\nonumber\\
&\
- \sum_{m=0}^{l-1}\pi_1\bd{B}_{m+2}(s,t)\Big(
D_{m+1}(s,t)
\wedge \pi_1E(s,t)\Big)\pi_{l+2}\,,
\label{rec rho}
\end{align}
with
\begin{equation}
 E(s,t)\ =\ \Delta(s,t)\bd{\rho}(s,t)
+Q^+\Xi\pi^1\bd{\Sigma}(s,t)\,,
\end{equation}
for $0\le l\le n-1$\,.\footnote{Note that the right hand side of (\ref{rec rho})
includes only $\bd{\Sigma}_{k+2}(s,t)$ and $\bd{\rho}_{k+2}(s,t)$ with
$0\le k< l$\,.}
Then, we can find that
\begin{align}
\pi_1\partial_t\bd{\Sigma}_{n+2}(s,t)\ =&\ [\bd{Q}\,,\pi_1\bd{\rho}_{n+2}(s,t)]\,,
\end{align}
with $\bd{\rho}_{n+2}(s,t)$ obtained by setting $l=n$ in Eq.~(\ref{rec rho}).
Hence, from the principle of mathematical induction we conclude
that $\partial_t\Sigma(s,t)=[\bd{Q},\bd{\rho}(s,t)]$ with
$\bd{\rho}(s,t)$ recursively determined by the equation
\begin{align}
 \ \pi_1\bd{\rho}(s,t)
=&\ 
\sum_{m=0}^\infty\pi_1\bd{\lambda}_{m+2}(s,t)\big(
D_{m+2}(s,t)\big)P_{n+2}\pi_{n+2}
\nonumber\\
&\
- \sum_{m=0}^\infty\pi_1\bd{B}_{m+2}(s,t)\Big(
D_{m+1}(s,t)\wedge \pi_1 E(s,t)\Big)P_{n+2}\pi_{n+2}\,.
\label{rec app}
\end{align}

Similarly, for $\bd{\Sigma}_2(s,t)$
and $\bd{\Sigma}_3(s,t)$\,, we have
\begin{equation}
\pi_1\partial_s\bd{\Sigma}_2(s,t)\ =\ 
[\bd{\eta},\pi_1\bd{\rho}_2(s,t)]\,,\qquad
\pi_1\partial_s\bd{\Sigma}_3(s,t)\ 
=\ [\bd{\eta},\pi_1\bd{\rho}_3(s,t)]\,.
\end{equation}
If we assume that
\begin{align}
&\ \pi_1\partial_s\bd{\Sigma}_{l+2}(s,t)\ =\ [\bd{\eta},\pi_1\bd{\rho}_{l+2}(s,t)]\,,\quad
\textrm{for}\quad 0\le l\le n-1\,,
\end{align}
then, we can show that
\begin{align}
\pi_1\partial_s\bd{\Sigma}_{n+2}(s,t)\ =\ 
[\bd{\eta},\pi_1\bd{\rho}_{n+2}(s,t)]\,.
\end{align}
Hence, it is proven from the principle of mathematical induction
that $\pi_1\partial_s\bd{\Sigma}(s,t)=[\bd{\eta},\pi_1\bd{\rho}(s,t)]$\,.

\section{Relation to the Erler-Okawa-Takezaki open superstring field theory}\label{open product}\label{app der}

When starting from the cubic theory, 
$\bd{M}_B^{(0)}(s)=\bd{m}_2|^0+s\,\bd{m}_2|^2$\,,
it is easy to see that that the generating functions $\bd{A}(s,t)$ 
and $\bd{\mu}(s,t)$ of the open superstring product and gauge product
have the restricted form
\begin{align}
 \bd{A}(s,t)\ =&\ \bd{A}(t) + s \bd{\mathcal{A}}(t)
\nonumber\\
=&\ \bd{A}|^0(t) + \bd{A}|^2(t) + s\,\bd{\mathcal{A}}|^0(t)\,,\\
 \bd{\mu}(s,t)\ =&\ \bd{\mu}|^0(t)\,,
\end{align}
where
\begin{align}
 \bd{A}|^0(t)\ =&\ \sum_{n=0}^\infty t^{n+1}\bd{A}^{(n+1)}_{n+2}|^0\,,\quad
 \bd{A}|^2(t)\ =\ \sum_{n=0}^\infty t^n\bd{A}^{(n)}_{n+2}|^2\,,
\nonumber\\
 \bd{\mathcal{A}}|^0(t)\ =&\ \sum_{n=0}^\infty t^n\bd{\mathcal{A}}^{(n)}_{n+2}|^0\,,\\
\bd{\mu}|^0(t)\ =&\ \sum_{n=0}^\infty t^n\bd{\mu}^{(n+1)}_{n+2}|^0\,.
\end{align}
%
Then, $[\bd{A}(s,t),\bd{\mu}(s,t)]^2\equiv0$
since the gauge product $\bd{\mu}(s,t)$ is only nonvanishing in the NS sector.
The differential equations (\ref{diff eq open}) are now simplified 
and decomposed to the four equations
\begin{subequations}\label{diff eq split}
\begin{align}
 \partial_t\bd{A}|^0(t)\ =&\ [\bd{Q},\bd{\mu}|^0(t)] + [\bd{A}|^0(t),\bd{\mu}|^0(t)]\,,\\
 \partial_t\bd{A}|^2(t)\ =&\ [\bd{A}|^2(t),\bd{\mu}|^0(t)]\,,\\
 \partial_t\bd{\mathcal{A}}|^0(t)\ =&\ [\bd{\mathcal{A}}|^0(t),\bd{\mu}|^0(t)]\,,\\
[\bd{\eta},\bd{\mu}|^0(t)]\ =&\ \bd{\mathcal{A}}|^0(t)\,,
\label{diff eq split 4}
\end{align}
\end{subequations}
satisfying the initial conditions $\bd{A}(0)=\bd{m}_2|^2$ 
and $\bd{\mathcal{A}}(0)=\bd{m}_2|^0$\,.
These equations are slight modification of those proposed in Ref.~\cite{Erler:2016ybs}.
In the previous method, almost the same equations that respect the Ramond number, instead of the cyclic
Ramond number, directly provide (the generating function of) the products with
$A_\infty$ structure. In our method, on the other hand, 
Eqs.~(\ref{diff eq split}) provide an intermediate products, which have to be transformed 
to the final form by the cohomomorphism. 
It can be seen, however, that the final products are the same.

In the present construction, we have an intermediate 
cyclic $A_\infty$ structure $\bd{Q}-\bd{\eta}+\bd{A}(t)$ 
which split into two (anti-)commutative
$A_\infty$ structures
\begin{align}
 \bd{D}(t)\ =&\ \bd{Q}+\bd{A}|^0_0(t)+\bd{A}|^2_2(t)\,,\\
 \bd{C}(t)\ =&\ \bd{\eta}-\bd{A}|^2_0(t)\,. 
\end{align}
Using the fact that $\bd{A}(t)$ satisfies the differential equations (\ref{diff eq split}), 
they can be rewritten as
\begin{align}
 \bd{D}(t)\ =&\ \hat{\bd{g}}(t)^{-1}(\bd{Q}+\bd{m}_2|_2)\hat{\bd{g}}(t)\,,\\
 \bd{C}(t)\ =&\ \hat{\bd{g}}(t)^{-1}(\bd{\eta}-\bd{m}_2|_0)\hat{\bd{g}}(t)\,,
\end{align}
where the cohomomorphism $\hat{\bd{g}}(t)$ is given by the path-ordered 
exponential of $\bd{\mu}|^0(t)$\,:
\begin{equation}
\hat{\bd{g}}(t)\ =\ \vec{\mathcal{P}}\exp\left[\int^t_0dt'\bd{\mu}|^0(t')\right]\,.
\end{equation}
A relation
\begin{equation}
 \hat{\bd{g}}(t)\bd{\eta}\hat{\bd{g}}(t)^{-1}\ =\ 
\bd{\eta}-\bd{m}_2|^0\,,
\end{equation}
follows from the Eq.~(\ref{diff eq split 4}) are also used.
The cohomomorphism 
$\pi_1\hat{\bd{F}}(t)^{-1}=\pi_1\id-\Xi\pi_1^1\bd{A}(t)$ transforming
$\bd{D}(t)$ to the final $A_\infty$ structure $\bd{M}(t)$\,,
is similarly written as
\begin{equation}
\hat{\bd{F}}(t)^{-1}\
=\  \hat{\bd{g}}(t)^{-1}(\hat{\bd{F}}^{-1})\hat{\bd{g}}(t)\,,\quad
\pi_1\hat{\bd{F}}^{-1}\ =\
\pi_1\id -\Xi\pi_1^1\bd{m}_2|^2_0\,,
\end{equation}
and thus, we find that
\begin{alignat}{3}
 \bd{M}(t)\ 
=&\ \hat{\bd{F}}(t)^{-1}\bd{D}(t)\hat{\bd{F}}(t)\
=&\ (\hat{\bd{F}}\hat{\bd{g}}(t))^{-1}(\bd{Q}+\bd{m}_2|_2)\hat{\bd{F}}\hat{\bd{g}}(t)\,,
\\
 \bd{\eta}\ =&\ \hat{\bd{F}}(t)^{-1}\bd{C}(t)\hat{\bd{F}}(t)\
=&\ (\hat{\bd{F}}\hat{\bd{g}}(t))^{-1}(\bd{\eta}-\bd{m}_2|_0)\hat{\bd{F}}\hat{\bd{g}}(t)\,.
\end{alignat}
It was proven in Ref.~\cite{Erler:2017onq} that 
the $A_\infty$ structure $\bd{M}=\bd{M}(1)$ 
agrees with that given in Ref.~\cite{Erler:2016ybs}.

\section{Cyclicity of generalized $A_\infty$ structure}\label{app cyclicity}

In this appendix, we show that the $A_\infty$ algebra (\ref{A infty}) is
cyclic with respect to $\Omega$\,. First, we note that $\bd{A}$ is cyclic with 
respect to $\omega_l$ by construction \cite{Erler:2013xta}:
\begin{equation}
 \langle\omega_l|(\pi_1\bd{A}\otimes\pi_1+\pi_1\otimes\pi_1\bd{A})\ =\ 0\,.
\end{equation}
Then, we can show that $\bd{a}$ is also cyclic with respect to $\omega_l$ as
\begin{align}
 \langle\omega_l|(\pi_1\bd{a}\otimes\pi_1+\pi_1\otimes\pi_1\bd{a})\ =&\
 \langle\omega_l|(\pi_1\bd{A}\hat{\bd{F}}\otimes\pi_1
+\pi_1\otimes\pi_1\bd{A}\hat{\bd{F}})
\nonumber\\
=&\
 \langle\omega_l|\big(\pi_1\bd{A}\hat{\bd{F}}\otimes
(\pi_1\hat{\bd{F}}-\Xi\pi_1^1\bd{A}\hat{\bd{F}})
+(\pi_1\hat{\bd{F}}-\Xi\pi_1^1\bd{A}\hat{\bd{F}})
\otimes\pi_1\bd{A}\hat{\bd{F}}\big)
\nonumber\\
=&\
\langle\omega_l|\big(
(\pi_1\bd{A}\otimes\pi_1
+\pi_1\otimes\pi_1\bd{A})(\hat{\bd{F}}\otimes\hat{\bd{F}})
\nonumber\\
&\hspace{1cm}
+ (\id\otimes\Xi-\Xi\otimes\id)(\pi_1\bd{A}\hat{\bd{F}}\otimes\pi_1\bd{A}\hat{\bd{F}})
\big)
\nonumber\\
=&\ 0\,,
\end{align}
where we used the relation
\begin{equation}
 \pi_1\id\ =\ \pi_1\hat{\bd{F}}-\Xi\pi_1^1\bd{A}\hat{\bd{F}}\,,
\end{equation}
following from the definition of $\hat{\bd{F}}^{-1}$ (\ref{open F}), 
and the fact that $\Xi$ is a BPZ even: 
\begin{equation}
 \langle\omega_l|(\id\otimes\Xi-\Xi\otimes\id)\ =\ 0\,.
\end{equation}
If we note that $[\bd{\eta},\bd{a}]=0$ it is easy to see that
$\bd{a}$ is also cyclic with respect to $\omega_s$\,:
\begin{align}
 \langle\omega_s|(\pi_1\bd{a}\otimes\id+\id\otimes\pi_1\bd{a})\
=&\ \langle\omega_l|(\xi_0\otimes\id)
(\pi_1\bd{a}\otimes\id+\id\otimes\pi_1\bd{a})
\nonumber\\
=&\ \sum_{n=0}^\infty
\langle\omega_l|(\xi_0\otimes\id)
(\pi_1\bd{a}\otimes\id+\id\otimes\pi_1\bd{a})
\pi_{n+3}([\eta\,,\xi_0]\otimes\id_{n+2})
\nonumber\\
=&\ - \sum_{n=0}^\infty
\langle\omega_l|(\pi_1\bd{a}\otimes\id+\id\otimes\pi_1\bd{a})
\pi_{n+3}(\xi_0\otimes\id_{n+2})\ 
\nonumber\\
=&\ 0\,,
\end{align}
on $\mathcal{H}_s$\,.
Since the BRST operator satisfies
\begin{equation}
\langle\Omega|(Q\otimes\id+\id\otimes Q)\ =\ 0\,,  
\end{equation}
on $\mathcal{H}^{res}$\,, we find that
\begin{align}
 \langle\Omega|(\pi_1\bd{M}\otimes\id+\id\otimes\pi_1\bd{M})\
=&\ \langle\omega_s|(\pi_1\bd{a}\otimes\id+\id\otimes\pi_1\bd{a})\
=\ 0\,,
\end{align}
on $\mathcal{H}^{res}$\,.

\vspace{1cm}


\end{document}